\documentclass[12pt]{article}
\usepackage{amsfonts,amssymb,amsmath}
\usepackage{theorem,cite}
\usepackage{color}
\newtheorem{definition}{Definition}[section]

\newtheorem{proposition}[definition]{Proposition}

\newtheorem{theorem}[definition]{Theorem}
\newtheorem{corollary}[definition]{Corollary}

\newtheorem{rmk}{Remark}[section]

\numberwithin{equation}{section}

\definecolor{brique}{rgb}{.9,.2,0}
\definecolor{blvert}{rgb}{0,.8,.85}
\definecolor{vertcl}{rgb}{0,1,.7}
\newcommand\vertcl[1]{\textcolor{vertcl}{#1}}
\newcommand\blvert[1]{\textcolor{blvert}{#1}}
\newcommand\brique[1]{\textcolor{brique}{#1}}

\def\lapth{
\begin{picture}(164,70)(0,-15)\thicklines
\put(0,0){\vertcl{\rule{20pt}{4pt}}}
\put(19,1){\vertcl{\line(1,3){23}}} 
\put(20,1){\vertcl{\line(1,3){23}}} 
\put(21,1){\vertcl{\line(1,3){23}}}
\put(22,1){\vertcl{\line(1,3){23}}}
\put(45,70){\vertcl{\line(1,-3){23}}} 
\put(44,70){\vertcl{\line(1,-3){23}}} 
\put(43,70){\vertcl{\line(1,-3){23}}}
\put(42,70){\vertcl{\line(1,-3){23}}}
\put(2,24){\vertcl{\rule{120pt}{4pt}}}
\put(65,0){\vertcl{\rule{60pt}{4pt}}}
\put(5,37){\Huge{\brique{\textbf{L}}}} 
\put(62,37){\Huge{\brique{\textbf{PTh}}}}
\put(12,-8){\blvert{\rule{92pt}{3.5pt}}}
\put(24,-15){\blvert{\rule{57pt}{3.5pt}}}
\put(36,-22){\blvert{\rule{30pt}{3.5pt}}}
\end{picture}
\raisebox{35pt}{
\begin{minipage}{320pt}\begin{center}
\textbf{Laboratoire d'Annecy-leVieux de Physique
Th\'eorique}\\[4ex]
website: \texttt{http://lappweb.in2p3.fr/lapth-2005/}
\end{center}
\end{minipage}}\\
\vspace{10pt}\quad \hrulefill\\
\vspace{10pt}}

\newcommand{\nonu}{\nonumber \\}

\newcommand{\hs}[1]{\hspace{#1 mm}}

\newcommand{\la}{\lambda}

   \def\cB{{\cal B}}   
   \def\cE{{\cal E}}

\def\cM{{\cal M}}   \def\cN{{\cal N}}   
      
   \def\cT{{\cal T}}   \def\cU{{\cal U}}
      
\def\cY{{\cal Y}}   

\def\fB{{\mathfrak B}}

\def\fM{{\mathfrak M}}

\def\fe{{\mathfrak e}}

\def\fm{{\mathfrak m}}

\newcommand{\CC}{{\mathbb C}}
\newcommand{\EE}{{\mathbb E}}
\newcommand{\II}{{\mathbb I}}

\newcommand{\NN}{{\mathbb N}}

\newcommand{\ZZ}{{\mathbb Z}}

\newcommand{\und}[1]{\underline{#1}}
\newcommand{\wh}[1]{\widehat{#1}}
\newcommand{\wt}[1]{\widetilde{#1}}
\newcommand{\mb}[1]{\hs{4}\mbox{#1}\hs{4}}

\newcommand{\prf}{\underline{Proof:}\ }
\newcommand{\finprf}{\null \hfill {\rule{5pt}{5pt}}\\[2.1ex]\indent}

\newcommand{\atopn}[2]{\genfrac{}{}{0pt}{}{#1}{#2}}

\begin{document}

\pagestyle{empty}
\markright{\today\dotfill DRAFT\dotfill }
\renewcommand{\thefootnote}{\fnsymbol{footnote}}
\newpage
\setcounter{page}{0}
\hspace{-1cm}\lapth

\vfill\vfill

\begin{center}
{\LARGE  {\sffamily 
Analytical Bethe Ansatz for closed and open $gl(\cM|\cN)$\\[1.2ex] 
super-spin chains in arbitrary representations\\[1.2ex]  
and for any Dynkin diagram}}\\

\vspace{10mm}
  
{\large E. Ragoucy$^{a}$\footnote{ragoucy@lapp.in2p3.fr}
and G. Satta$^{a\,b}$\footnote{satta@fis.uniroma3.it}\\[.42cm]
$^{a}$ \textsl{Laboratoire de Physique Th{\'e}orique LAPTH\footnote{UMR 5108 
du CNRS, associ{\'e}e {\`a} l'Universit{\'e} de
Savoie.}\\[.242cm]
LAPP, BP 110, F-74941  Annecy-le-Vieux Cedex, France. \\[.40cm]
$^{b}$ Dipartimento di Fisica, Universit{\`a} di Roma 3 and INFN
\\[.242cm]
Via Vasca Navale 84, I-00146 Roma, Italy.} }

\end{center}
\vfill\vfill

\begin{abstract}
We present the analytical Bethe ansatz for spin chains based on the 
superalgebras $gl(\cM|\cN)$, $\cM\neq\cN$,
with at each site an arbitrary 
representation (and including inhomogeneities).
 The calculation is done 
for closed and open spin chains. In this latter case, the boundary 
matrices $K_{\pm}(\lambda)$ are of general type, 
provided they commute. 
We compute the Bethe ansatz equations in full generality, and for any 
type of Dynkin diagram. 
Examples are worked out to illustrate the techniques.
\end{abstract}

\vfill

\begin{center}
\end{center}

\vfill

\rightline{LAPTH-1189/07}
\rightline{\texttt{hep-th/0706.3327}}
\rightline{June 2007}

\newpage
\pagestyle{plain}
\renewcommand{\thefootnote}{\arabic{footnote}}
\setcounter{footnote}{0}

\section{Introduction}
The possibility of constructing and solving by
algebraic and/or analytical methods one-dimen\-sional interacting quantum
spin chains, is one of the major achievements of quantum
integrable systems. It allows the determination of the spectrum, 
eigenvectors and (at least partially) the calculation of correlation 
functions. The main tool is the quantum $R$-matrix, obeying a
cubic Yang-Baxter equation, the ``coproduct'' properties of which allow the
building of a periodic $L$-site transfer matrix with identical exchange
relations and the subsequent derivation of
quantum commuting Hamiltonians \cite{baxter}. A similar structure 
arises for non-periodic (open) spin chains. These are characterised by a
second object: the reflection matrix $K$, obeying a quadratic consistency
equation with the $R$ matrix
\cite{cherednik,sklyanin,KuSk,KuSa,DoKuMu}. Using again ``coproduct-like''
properties of this structure one 
constructs suitable transfer matrices yielding (local) commuting spin chain
Hamiltonians by combining $K$ and semi-tensor products of $R$
\cite{sklyanin}. 

Recently, a more algebraic approach to the analytical Bethe 
ansatz has been developped, allowing a `universal' approach (i.e; 
whatever the spins on the chain) to the spectrum of the 
transfer matrix, and the corresponding Bethe equations.  This 
framework has been developped for open and closed spin chains, based 
on $gl(\cN)$ \cite{byebye} and $\cU_{q}(gl_{N})$ 
\cite{qbound} algebras. 

On an other hand, 
quantum supersymmetric integrable systems appeared \cite{Minahan}
in the context of $N=4$ super-Yang-Mills (SYM) theories, in the loop
expansion of the dilatation operator, used for the computation of
anomalous dimensions of trace operators. In fact, it seems that (at
least for the first loop corrections) that the dilatation operator can
be identified with some super-spin chain Hamiltonian, the type of the 
chain depending both on the (sub)sector of the SYM theory one
considers, and on the order of loop correction, see e.g. \cite{BDS}.  

Hence, it is the right time to give a general overview of 
the possible integrable closed and open super-spin chains that one can construct 
starting from a $gl(\cM|\cN)$ superalgebra and arbitrary 
spins on the chain. We will study the spectrum and Bethe equations 
associated to these chains.
Closed spin chains  based on $sl(\cM|\cN)$ superalgebras in the
distinguished diagram were
studied in \cite{Kulish} and \cite{saleur} and, in the case of alternating
fundamental-conjugate  
representations of $sl(\cM|\cN)$ in \cite{mmartins}. 
In 
\cite{tsuboi}, closed spin chains in the fundamental representation but for 
any type of Dynkin diagram where studied using the Baxter 
$Q$-operator, and generalized in \cite{tsuboi2} to a chain where all 
the spins are in a (type 1) typical representation depending on a free 
parameter. General approach using Hirota equation was done in \cite{zaka}.
Open spin chains based on $sl(1|2)$ have been studied in details in
e.g. \cite{GR,Essler}. The $sl(\cM|\cN)$ case with spins in the
fundamental representation, with diagonal $K(u)$ matrices,
but for any type of Dynkin diagrams have been done in
\cite{selene}. The deformed case for fundamental
representations but general $K(u)$ matrices have been studied in
\cite{HouFanYue}.
We will use the algebro-analytical 
framework developped in \cite{byebye,qbound}, applied to superalgebras. It will 
provide a `universal' presentation for all chains (whatever the 
representations that enter the chain), for closed and open cases.
A particularity of superalgebras 
(that do not share usual algebras) 
is the existence of different Dynkin diagrams for the same superalgebra. 
This leads to different presentations of the spectrum of the same transfer 
matrix, hence to different Bethe equations: the presentation is also universal 
in the sense that it applies  for 
all Dynkin diagramms of the superalgebra.

The plan of the paper is as follows. In section \ref{sec:alg}, we 
present the algebraic structures that are needed for the construction 
of super-spin chains: the super-Yangian based on $gl(\cM|\cN)$ for 
closed chains and the reflection superalgebra for open chains. Then, 
in section \ref{sec:closed}, we construct the closed spin chains, give 
their spectrum and their Bethe equations, in the case of 
distinguished Dynkin diagram. Section 
\ref{sec:arbitraryDD} is devoted to the 
general form of the Bethe equations for each of the different 
Dynkin diagramms 
of the superalgebra.
The case of open super-spin chain is treated in section 
\ref{sec:open}, including the different presentations associated to 
different Dynkin diagrams. Finally, section \ref{sec:examples} 
illustrates our method on examples.
 
\section{Algebraic structures\label{sec:alg}}
\subsection{Graded spaces}
We will work on $\ZZ_{2}$-graded spaces $\CC^{\cM\vert \cN}$, with
$\ZZ_{2}$-grade
\begin{equation}
[\ ]:\left\{ \begin{array}{lcl}  \NN_{\cM+\cN} & \to & \{0,1\} \\
                               j & \mapsto & [j]
  \end{array}\right.
\end{equation}
where $\mathbb N_{\cM+\cN} = \left\{1,2,...,\cM+\cN\right\}$.
The elementary $\CC^{\cM\vert \cN}$ vectors $e_{i}$ and
$End(\CC^{\cM\vert
\cN})$ matrices $E_{ij}$ have grade 
\begin{equation}
[e_{i}]=[i] \mb{and} [E_{ij}]=[i]+[j]. 
\end{equation}
The tensor product is graded accordingly:
\begin{equation}
(E_{ij} \otimes E_{kl})(E_{ab} \otimes E_{cd}) =
(-1)^{([k]+[l])([a]+[b])}(E_{ij}E_{ab}\otimes E_{kl}E_{cd})\,.
\end{equation}
The permutation operator
\begin{equation}
P_{12} = \sum_{i,j=1}^{\cM+\cN} (-1)^{[j]}E_{ij}\otimes E_{ji}
\end{equation}
is also graded
\begin{eqnarray}
P_{12} (e_i \otimes e_j) = (-1)^{[i][j]}\,e_j \otimes e_i
\mb{and} P_{12}(E_{ij} \otimes E_{kl})P_{12}=
(-1)^{([i]+[j])([k]+[l])}\,E_{kl}\otimes
E_{ij}\,.
\end{eqnarray}
The permutation operator obey the relation $P_{12}^2=\II\otimes\II$,
so that it is symmetric:
\begin{equation}
 P_{21}=P_{12}\,P_{12}\,P_{12}=P_{12}
\end{equation}
Together with the $\ZZ_{2}$-grading, we will use a graded commutator 
$[.,.\}$, which is graded antisymmetric and obeys a graded Jacobi 
identity.

Unless explicitly specified, we will work with the
\textit{distinguished}
$\ZZ_{2}$-grade defined by
\begin{equation}
[i]=\left\{ \begin{array}{ll}  0\,, & 1 \leq i \leq \cM\,,\\
                               1\,, & \cM+1\leq i \leq \cM+\cN\,.
  \end{array}\right.
\end{equation}
However, in some cases, we will use different grading, such as the
\textit{symmetric} $\ZZ_{2}$-grade, defined for $\cN=2n$:
\begin{equation}\label{symmgrad}
[i]=\left\{ \begin{array}{ll}  0\,, & 1 \leq i \leq n \mb{and}
\cM+n+1\leq i\leq \cM+\cN\,,\\
                               1\,, & n+1\leq i \leq \cM+n\,.
  \end{array}\right.
\end{equation}
The name of these grading refers to the $gl(\cM|\cN)$ Dynkin diagram
(and simple roots) they are associated to, see below.

\subsection{The $gl(\cM|\cN)$ superalgebra}
The Lie superalgebra $gl(\cM|\cN)$  
is a  $\ZZ_2$-graded vector space over $\CC$ spanned by the basis 
$\{\cE_{ab}| a,b=1,2,...,\cM+\cN\}$. The gradation is defined
by the $\ZZ_{2}$-grade $[\ ]$ through:
\begin{equation}
 [\cE_{ab}]=[a]+[b]\,.
 \label{Z2grad}
\end{equation}
The bilinear graded commutator 
associated to $gl(\cM|\cN)$ is defined by: 
\begin{equation}\label{glMN}
{[}\cE_{ab},\,\cE_{cd}\}
=\delta_{cb}\,\cE_{ad}-(-1)^{([a]+[b])([c]+[d])}
\delta_{ad}\,\cE_{cb}\,.
\end{equation}
Gathering the generators $\cE_{ab}$ into a single matrix
\begin{equation}
\EE=\sum_{a,b=1}^{\cM+\cN} (-1)^{[a]}\,\cE_{ab}\,E_{ab}
\end{equation}
the above commutation relations can be recasted as
\begin{equation}
\Big[\EE_{1}\,,\,\EE_{2}\Big\} = P_{12}\Big(\EE_{2}-\EE_{1}\Big) 
\label{CRglMN}
\end{equation}
where $\EE_{1}=\EE\otimes\II$ and $\EE_{2}=\II\otimes\EE$.

Although the $gl(\cM|\cN)$ superalgebra is a graded version of the
$gl(\cM+\cN)$ algebra, they differ on several points, a common feature
when comparing Lie algebras and superalgebras, 
see e.g. \cite{dico} for more details. In particular,
there exist several inequivalent simple roots systems, leading to 
different presentations of the same superalgebra. One can relate 
these different systems to a choice of the $\ZZ_{2}$-grade.
To each inequivalent simple roots system correspond a Dynkin diagram, 
so that a superalgebra possesses several Dynkin diagram. Note however 
that any Dynkin diagram defines uniquely a superalgebra.

\subsection{The super-Yangian $\mathcal Y (\cM|\cN)$}
$\mathcal{Y} (\cM|\cN)$ is the graded unital associative algebra, with
generators $T^{(n)}_{ab}$, $n> 0$, $a,b = 1,...,\cM+\cN$, with
$\ZZ_{2}$-grade
\begin{equation}
[T^{(n)}_{ab}] = [a]+[b]\,, \qquad \forall\, a,b,n\,.
\end{equation}
We gather $\mathcal Y (\cM|\cN)$ generators in matrix
form with $T^{(0)}_{ab} = \delta_{ab}$
\begin{equation}
T(u) \doteq \sum_{a,b=1}^{\cM+\cN} \sum_{n\geq 0}
\frac{\hbar^n}{u^n}\,T^{(n)}_{ab}\,E_{ab} \doteq \sum_{n\geq 0} 
\frac{\hbar^n}{u^n}\,T^{(n)}
\doteq
\sum_{a,b}^{\cM+\cN} T_{ab}(u)\,E_{ab}\,,
\end{equation}
which is an even element of $ \mathcal Y (\cM|\cN)[u^{-1}]\otimes
End(\CC^{\cM\vert \cN})$. Here and below, the space
$End(\CC^{\cM\vert \cN})$
will be refered as the auxiliary space, while (the copies of) the
super-Yangian $\cY(\cM|\cN)[u^{-1}]$ will be called the quantum
space(s).

$\mathcal Y (\cM|\cN)$ commutation relations are given by the
so-called FRT exchange relation\cite{FRT}
\begin{equation}\label{rtt}
R_{12}(u-v) \,T_1(u)\,T_2(v) = T_2(v)\,T_1(u)\, R_{12}(u-v)\,,
\end{equation}
each side of the equation being an element of 
$\cY(\cM|\cN)[u^{-1}]\otimes End(\CC^{\cM\vert \cN})\otimes 
End(\CC^{\cM\vert \cN})$,
and where we have introduced the super-Yangian $R$-matrix\footnote{The
normalization is chosen in such a way that $R(u)$ is analytic in $u$.}
\begin{equation}\label{Rmatrix}
R_{12}(u)= u\,\II_{\cM+\cN}\otimes \II_{\cM+\cN}-{\hbar}P_{12}\,.
\end{equation}
It acts on the two auxiliary spaces associated to
$T_1(u)=T(u)\otimes\II_{\cM+\cN}$ and $T_2(u)=\II_{\cM+\cN}\otimes
T(u)$. 
The deformation parameter $\hbar$ is in fact irrelevant (provided it
is not zero), hence  it is in general set to 1 for algebraic studies.
However, in the context of spin chain models, it is set to $-i$, 
so that we keep it free to encompass these two choices.
Note that the $R$-matrix is a globally even one. Its inverse reads
\begin{equation}
R^{-1}_{12}(x) =
\frac{1}{x^2-\hbar^{2}}\,\left(x\,\II\otimes\II +
{\hbar}\,P_{12}\right)
=\frac{-1}{x^2-\hbar^{2}}\,R_{12}(-x)\,.
\end{equation}

Projecting the relation
(\ref{rtt}) on elementary matrices $E_{ab}\otimes E_{cd}$, one gets
\begin{equation}
\Big[ T_{ab}(u)\,, T_{cd}(v)\Big\} =
\frac{(-1)^{\eta(a,b,c)}\,\hbar}{u-v}\Big(T_{cb}(u)\, T_{ad}(v) -
T_{cb}(v)\,
T_{ad}(u)  \Big)\,,
\end{equation}
where $\eta(a,b,c) = [a]([b]+[c])+[b][c]$ and $
\left[\cdot\,,\cdot\right\} $ denotes the supercommutator. 

Expanding the commutation relation in $u^{-1}$ and $v^{-1}$, we obtain
\begin{equation}
\left[ T_{ab}^{(m)}\,, T_{cd}^{(n)}\right\} =
(-1)^{\eta(a,b,c)}\,\sum_{p=0}^{min(m,n)-1}\left(T_{cb}^{(p)}\,
T_{ad}^{(m+n-1-p)}- T_{cb}^{(m+n-1-p)}\,T_{ad}^{(p)} \right)\,,
\end{equation}
This commutation relation shows that the generators
$(-1)^{[a]}\,T_{ab}^{(1)}$
span a $gl(\cM|\cN)$ sub-superalgebra of the super-Yangian. 
Conversely, one can construct a morphism from the Lie superalgebra to
the super-Yangian, called the evaluation map:
\begin{equation}
ev:\ \left\{\begin{array}{lcl}
gl(\cM|\cN) &\to & \cY(\cM|\cN) \\[1.2ex]
T_{ab}(u) &\mapsto& \displaystyle
\delta_{ab}+\frac{\hbar}{u}\,(-1)^{[a]}\,\cE_{ba}\\[1.2ex]
\displaystyle T(u) &\mapsto& \displaystyle
\II+\frac{\hbar}{u}\,\EE
\end{array}\right.
\end{equation}
Using the commutation relations (\ref{CRglMN}) of $gl(\cM|\cN)$, it
is easy to show
that $ev(T(u))$ obey the relation (\ref{rtt}).

Two subalgebras of $\cY(\cM|\cN)$ will be used in the
following: the Yangian $\cY(\cM)$, generated by 
$\left\{  T_{ab}(u)\,, [a]=[b]=0\right\}$ 
and the Yangian $\cY_{-\hbar}(\cN)$, generated by 
$\left\{  T_{ab}(u)\,,[a]=[b]=1 \right\}$.
The generators of these subalgebras are obtained from $T(u)$
using  suitable $End(\CC^{\cM\vert\cN})$ projectors:
\begin{eqnarray*}
&&T^{(\cM)}(u) = \II_{\cM}\,T(u)\,\II_{\cM}
\mb{with} \II_{\cM} = \sum_{i, [i]=1}E_{ii}\,,
\\
&&T_{-\hbar}^{(\cN)}(u) = \II_{\cN}\,T(u)\,\II_{\cN}
\mb{with} \II_{\cN} = \sum_{i, [i]=0}E_{ii}\,.
\end{eqnarray*}

The map
\begin{equation}
\Delta : \left\{\begin{array}{lcl}
\mathcal Y (\cM|\cN)[u^{-1}] &\to &
\mathcal Y (\cM|\cN)[u^{-1}]\otimes \mathcal Y
(\cM|\cN)[u^{-1}]\\[1.2ex]
T_{ij}(u) &\mapsto& \displaystyle
\Delta\left(T_{ij}(u)\right) = \sum_{k=1}^{\cM+\cN}T_{ik}(u) \otimes
T_{kj}(u) 
\end{array}\right.
\end{equation}
is an homomorphism of $\mathcal Y(\cM|\cN)$. Gathering the generators 
into matrices, it rewrites 
\begin{equation}\label{coprod1}
\Delta\left(T(u)\right) = T(u)\dot \otimes
T(u)\in\mathcal Y (\cM|\cN)[u^{-1}]\otimes \mathcal Y
(\cM|\cN)[u^{-1}]\otimes End(\CC^{\cM\vert \cN})\,.
\end{equation}
$\Delta$ is coassociative:
\begin{equation}
\Delta^{(n)} = (\Delta^{(n-1)} \otimes \mathrm{id} )\, \Delta =
(\mathrm{id} \otimes \Delta^{(n-1)})\, \Delta\,.
\end{equation}

\subsubsection{Highest weight vectors and modules}

A $\mathcal Y(\cM|\cN)$ module $V$ 
is said to be highest weight if there exists $v\,\in\,V$ such that
\begin{equation}\label{triang}
\left\{ \begin{array}{ll}  T_{aa}(u) \,v =  \lambda_a(u)\,v\,, \quad
\lambda_a(u)\,\in\,  
1+u^{-1}\CC[u^{-1}] & \forall\, a=1,...,\cM+\cN \\[1.2ex]
T_{ab}(u) \,v = 0\,, & 1\leq b<a\leq\cM+\cN
 \end{array}\right.
\end{equation}
The vector $\lambda(u) \doteq
(\lambda_{1}(u),...,\lambda_{\cM+\cN}(u))$ is the highest weight of
$V$, and $v$ 
a highest weight vector.  
The following theorems have been proved in \cite{Zhang} 
\begin{theorem}
 Any finite--dimensional irreducible representation of $\mathcal
Y(\cM|\cN)$ admits
a unique highest weight vector (up to
normalization).
\end{theorem}

\begin{theorem}
 An irreducible representation with highest weight $\lambda(u)$ is
finite--dimensional if and only if
\begin{equation}\label{monic}
\frac{\lambda_a(u)}{\lambda_{a+1}(u)} =
\frac{P_a(u+\hbar)}{P_a(u)}\,,\quad 1\leq a<\cM+\cN \mb{and} a
\neq \cM\,, 
\qquad
\frac{\lambda_\cM(u)}{\lambda_{\cM+1}(u)} =
\frac{P_\cM(u)}{P_{\cM+\cN}(u)}\,,
\end{equation}
where all $P_a(u)$ are monic polynomials.
\end{theorem}

Among the finite-dimensional highest weight representations, 
there is a class of particular interest, constructed from the
evaluation map:
an evaluation representation $ev_{\pi_{\mu}}=\pi_{\mu}\,\circ\,ev$ 
is a morphism from the super-Yangian 
$Y(\cM|\cN)$ to a highest weight irreducible representation
$\pi_{\mu}$ of 
$gl(\cM|\cN)$.
The morphism is given by:
\begin{equation}\label{evaluation}
ev_{\pi_{\mu}}(T_{ij}(u))=\delta_{ij} + 
(-1)^{[i]}\,\pi_{\mu} (\cE_{ji})\,\frac{\hbar}{u-a}\quad 
\forall i,j\in\{1,...,\cM+\cN\}\,,\quad a\in \mathbb C\,,
\end{equation}
where the dependance (that will be left implicit in what follows) 
of $ev_{\pi_{\mu}}$ on an arbitrary complex shift
of the spectral parameter  has been introduced.
One has
\begin{equation}
ev_{\pi_{\mu}}(T_{ij}^{(1)})=(-1)^{[i]}\,\pi_{\mu} (\cE_{ji})
\ ;\ ev_{\pi_{\mu}}(T_{ij}^{(r)})=0\mb{for} r>1\,.
\end{equation}

The highest weight $\mu(u)=(\mu_1(u),...,\mu_{\cM+\cN}(u))$ of
the representation $ev_{\pi_{\mu}}$ is given by:
\begin{equation}
\mu_{i}(u) = 1+(-1)^{[i]}\,\mu_{i} \frac{\hbar}{u-a}\,,\ \forall i\in
\{1,...,\cM+\cN\}
\end{equation}
where $\mu = (\mu_{1},...,\mu_{\cM+\cN})$ is the highest weight of
$\pi_{\mu}$.
The evaluation morphism associated to the fundamental representation of $gl(\cM|\cN)$, with
highest weight $\mu_f=(1,0,...,0)$, provides the $R$ matrix (\ref{Rmatrix}).

\begin{theorem}\cite{Zhang}
Any finite-dimensional irreducible representation of $\cY(\cM|\cN)$
can be obtained through the
tensor products\footnote{Note however that one has sometimes to 
make a quotient to get an irreducible representation from these 
tensor products.} of such evaluation representations.
\end{theorem}
    
Let  $\{ev_{\pi_i}\}_{i=1,...,s}$ be a set of evaluation
representations. 
The tensor products of these $s$ representations 
$ev_{\vec{\pi}}=ev_{\pi_1}\otimes ...\otimes ev_{\pi_s}\ \circ\
\Delta^{(s)}$ is a morphism from $Y(\cM|\cN)$ to the tensor product 
of $gl(\cM|\cN)$ representations $\vec{\pi}=\otimes_{i}\pi_{i}$ given
by:
\begin{equation}\label{evaluationprod}
ev_{\vec{\pi}}(T_{ab}(u))=\sum_{i_{1},\ldots,i_{s-1}}
 ev_{\pi_{1}}(T_{ai_{1}}(u))\otimes
ev_{\pi_{2}}(T_{i_{1}i_{2}}(u))\otimes\cdots\otimes
ev_{\pi_{s}}(T_{i_{s-1}b}(u))
\end{equation}

\subsubsection{The generators $T^*(u)$}
For the study of superspin chains, we will need also 
\begin{equation}
T^{*}(u)=T^{-1}(u)^t=\sum_{a,b=1}^{\cM+\cN} T^{*}_{ab}(u)\,E_{ab}
\end{equation}
where the graded transposition is defined as
\begin{equation}
  \label{eq:st}
  A^{t} = \sum_{i,j=1}^{\cM+\cN} (-1)^{[i][j]+[j]} \; A_{ji} \,
E_{ij} = \sum_{i,j=1}^{\cM+\cN}
  \left(A^t\right)_{ij} \, E_{ij}\,,
\mb{that is}\left(A^t\right)_{ij}=(-1)^{[i][j]+[j]} \; A_{ji}\,.
\end{equation}

These generators have been introduced by Nazarov \cite{naza}, and
it is easy to see that they obey the same
relations as $T(u)$:
\begin{equation}
R_{12}(u-v) \,T^*_1(u)\,T^*_2(v) = T^*_2(v)\,T^*_1(u)\, R_{12}(u-v)
\,.
\label{rt*t*}
\end{equation}
Thus, the map 
\begin{equation}\label{isom}
\varphi\,: T(u) \mapsto  T^*(u)
\mb{i.e.} \varphi\left[T_{ij}(u)\right] =
T^*_{ij}(u)=(-1)^{[i][j]+[j]}T^{-1}_{ji}(u)
\end{equation} 
is an algebra isomorphism.
The exchange relation between $T^*(u)$ and $T(v)$ reads
\begin{eqnarray}
&& R^{t_{1}}_{12}(v-u)\,T_1^{*}(u)\,T_2(v) =
T_2(v)\,T_1^{*}(u)\, R^{t_{1}}_{12}(v-u) 
\,,\label{rt*t}\\
&& R^{t_{2}}_{12}(v-u+\hbar(\cM-\cN))\,T_1(u)\,T^{*}_2(v) =   
T^{*}_2(v)\,T_1(u)\, R^{t_{2}}_{12}(v-u+\hbar(\cM-\cN))
\label{rtt*}\,,
\end{eqnarray}
where the superscript $t_{1}$ (resp. $t_{2}$) denotes the 
transposition in the auxiliary space 1 (resp. 2). We have used the 
inversion formula
\begin{equation}
R^{t_{2}}_{12}(x)^{-1}=\frac{-1}{x(x-\hbar)}\,R^{t_{2}}_{12}(\hbar(\cM-\cN)-x)\,.
\end{equation}
One has also
\begin{eqnarray}
\left[ T^*_{nm}(u)\,,T_{kl}(v)   \right\} &=&
\frac{\hbar\,(-1)^{[k][m]}}{u-v}\Big(
\delta_{ml}(-1)^{[m]+[k][n]}
\sum_{a=1}^{\cM+\cN}(-1)^{[a][n]+[a]}T_{ka}(v)T^*_{na}(u)\nonu
&&\qquad\qquad -\delta_{nk}(-1)^{[n]}
\sum_{a=1}^{\cM+\cN}(-1)^{[a][m]}T^*_{am}(u)T_{al}(v)\Big)\,.
\label{scr4}
\end{eqnarray}

\subsubsection{Liouville contraction and crossing symmetry}
 The starting point is the equality
\begin{equation}
R^{t_{2}}_{12}(0)=\hbar\, Q_{12}=\hbar\, P_{12}^{t_{2}}
=\hbar\,  \sum_{i,j=1}^{\cM+\cN}
(-1)^{[j]+[i]+[i][j]}E_{ij}\otimes E_{ij}\,.
\end{equation}
When $\cM \neq \cN$, $Q_{12}$ is (up to normalization) a
one-dimensional projector
$Q_{12}^2 = (\cM-\cN)Q_{12}$ of 
$\mathrm{End}(\mathbb C^{(\cM|\cN)})$. Remark that it is \und{not}
symmetric:
\begin{equation}
Q_{21}= P_{12}\,Q_{12}\,P_{12}=P_{12}^{t_{1}}
= \sum_{i,j=1}^{\cM+\cN} (-1)^{[i][j]}E_{ij}\otimes E_{ij}\neq
Q_{12}=P_{12}^{t_{2}}\,.
\end{equation}

Then, from (\ref{rtt*}), one
proves that there exist a central element $Z(u)$ of
$\mathcal Y(\cM|\cN)$ such that:
\begin{equation}
\label{TTQ}
  Q_{12}\, T_1(u+\hbar(\cM-\cN))\, T_2^*(u) = T_2^*(u)\,
T_1(u+\hbar(\cM-\cN)) \, Q_{12}
	= Z(u)\, Q_{12}\,.
\end{equation}
We refer to the original work \cite{naza} for more details.

Remark that this relation induces a crossing relation for the
super-Yangian generators. Indeed, starting from (\ref{TTQ}), one gets 
\begin{equation}
 Q_{12}\, T_1(u+\hbar(\cM-\cN)) = Z(u) \, Q_{12}\, T_2^*(u)^{-1}
\end{equation}
which, upon transposition in space 2 and multiplication by $P_{12}$,
leads to
\begin{equation}\label{primarel}
\left(\left(T^{-1}(u)^t\right)^{-1}\right)^{t} =
\frac{1}{Z(u)}\,T(u+\hbar(\cM-\cN))\,,
\end{equation}
or analogously
\begin{equation}\label{HAT}
T^t(u)^{-1} = \frac{1}{Z(u-\hbar(\cM-\cN))}\,
T^{-1}(u-\hbar(\cM-\cN))^{t}\,.
\end{equation}
This relation is nothing but the crossing symmetry for the $R$-matrix,
but extended at the super-Yangian (abstract) level. It 
 allows a crossing relation for the transfer matrix (see
below).\\
Note that this
calculation is also valid for the `usual' Yangian $\cY(\cN)$. 
In particular, for the $\cY(\cM)$ and $ \cY_{-\hbar}(\cN)$ subalgebras of $\cY(\cM|\cN)$ one has
\begin{eqnarray}
\Big( T^{(\cN)}_{-\hbar}(u)^t\Big)^{-1} &=&  z^{(\cN)}_{-\hbar}(u)\,
\Big(T^{(\cN)}_{-\hbar}(u+\hbar\cN)^{-1}\Big)^t\,,
\label{submatrix}
\\
\Big( T^{(\cM)}(u)^t\Big)^{-1} &=&  
z^{(\cM)}(u)\,\Big(T^{(\cM)}(u-\hbar\cM)^{-1}\Big)^t
\end{eqnarray}
for some scalar functions $z^{(\cM)}(u)$ and $z^{(\cN)}_{-\hbar}(u)$.
They are related to the quantum 
determinant of $\cY(\cM)$ (see e.g. \cite{Molev2001}) through:
\begin{equation}\label{rapporto}
z^{(\cM)}(u) =
\frac{\mathrm{qdet}\,T^{(\cM)}(u-\hbar)}{\mathrm{qdet}\,
T^{(\cM)}(u)}\,.
\end{equation}
We remind that the quantum determinent
$\mathrm{qdet}\,T(u)$ is the central element of $\cY(\cM)$ 
given by
\begin{eqnarray}
\mathrm{qdet}\,T(u) &=& \sum_{\sigma\in S_{\cM}} \mathrm{sgn}(\sigma)
\, T_{\sigma(1)1}(u)\cdots T_{\sigma(\cM)\cM}(u-\hbar(\cM-1)) \\
&=& \sum_{\sigma\in S_{\cM}} \mathrm{sgn}(\sigma)
\, T_{1\sigma(1)}(u-\hbar(\cM-1))\cdots T_{\cM\sigma(\cM)}(u)\,.
\end{eqnarray}
Its value in the highest weight representation
is computed through application of the above formula on $v^{+}$. 
For the Yangians $\cY(\cM)$ and $\cY_{-\hbar}(\cN)$ in $\cY(\cM|\cN)$,
we get:
\begin{eqnarray}
\mathrm{qdet}\,T^{(\cM)}(u) &=& 
\lambda_1(u-\hbar(\cM-1)) \cdots \lambda_{\cM}(u)\,,\\
\mathrm{qdet}\,T^{(\cN)}_{-\hbar}(u) &=& 
\lambda^{(\cN)}_1(u+\hbar(\cN-1)) \cdots \lambda^{(\cN)}_{\cN}(u)\,,
\end{eqnarray}
where $\lambda^{(\cN)}_j(u)=\lambda_{\cM+j}(u)$, $j=1,\ldots,\cN$.
It leads to the following expressions:
\begin{eqnarray}
z^{(\cM)}(u) &=& \frac{\lambda_1(u-\hbar\cM) \cdots \lambda_{\cM}(u-\hbar)}
{\lambda_1(u-\hbar(\cM-1)) \cdots \lambda_{\cM}(u)}\,,
\\
z^{(\cN)}_{-\hbar}(u) &=& 
\frac{\mathrm{qdet}\,T^{(\cN)}_{-\hbar}(u+\hbar)}
{\mathrm{qdet}\,T^{(\cN)}_{-\hbar}(u)}
= 
\frac{\lambda^{(\cN)}_1(u+\hbar\cN) \cdots \lambda^{(\cN)}_{\cN}(u+\hbar)}
{\lambda^{(\cN)}_1(u+\hbar(\cN-1)) \cdots
\lambda^{(\cN)}_{\cN}(u)}\,.
\label{eq:liouv-N}
\end{eqnarray}
The calculation of the function $Z(u)$ needs the use of the quantum 
Berezinian, see section \ref{qber}.

\subsubsection{Relations for $T^{-1}(u)$}
We will need the commutation relations for the inverse of $T(u)$,
defined by the relation
\begin{equation}
    T(u)\,T^{-1}(u)=\II \mb{with} 
    T^{-1}(u) = \sum_{a,b=1}^{\cM+\cN} T'_{ab}(u)\,E_{ab}\,,\
T'_{ab}(u)=\delta_{ab}+\sum_{n>0}\,\left(\frac{\hbar}{u}\right)^{n}\,T_{ab}^{'(n)}\,.
\end{equation}
This relation is understood as a series in $u^{-1}$, so that expanding
the above equality, one can
reconstruct the generators $T_{ab}^{'(n)}$ from the generators
$T_{ab}^{(n)}$, according to
\begin{equation}
T_{ab}^{'(n)}=-T_{ab}^{(n)}-\sum_{c=1}^{\cM+\cN} 
\sum_{p=1}^{n-1}T_{ac}^{'(n-p)}T_{cb}^{(p)}\,.
\label{expr:t'}
\end{equation}
{From} the relation (\ref{rtt}), one deduces that
\begin{eqnarray}
&&T_2(v)\, R_{12}(v-u)\,T_1^{-1}(u) =   T_1^{-1}(u)\,
R_{12}(v-u)\,T_2(v)\,,\\
&&T_2^{-1}(v)\, R_{12}(v-u)\,T_1(u) =   T_1(u)\,
R_{12}(v-u)\,T_2^{-1}(v)\,,\\
&& R_{12}(v-u)\,T_1^{-1}(u)\,T_2^{-1}(v) =
T_2^{-1}(v)\,T_1^{-1}(u)\,
R_{12}(v-u)\,,
\end{eqnarray}
which upon projection on $E_{mn}\otimes E_{kl}$ leads to
$$
\Big[ T'_{mn}(u)\,,T_{kl}(v)   \Big\} =
\frac{\hbar\,(-1)^{[k][n]}}{u-v}\,\sum_{a=1}^{\cM+\cN}\Big(
\delta_{ml}(-1)^{[k][m]+[m][n]}\,T_{ka}(v)\,T'_{an}(u)
-\delta_{nk}\,T'_{ma}(u)\,T_{al}(v)\Big)\,.
$$
Expanding in $u^{-1}$ and $v^{-1}$, one gets
\begin{equation}\label{tt'-serie}
\left[ T'^{(p+1)}_{mn}\,, T_{kl}^{(s)}   \right\} =
(-1)^{[k][n]}\sum_{r=0}^p\sum_{a=1}^{\cM+\cN}
\left(\delta_{ml}(-1)^{[k][m]+[m][n]}T_{ka}^{(s+r)}T_{an}'^{(p-r)}  -
\delta_{nk}T_{ma}'^{(p-r)}T_{al}^{(s+r)} \right)\,.
\end{equation}

\begin{proposition}
Let $v^+$ be a highest weight vector of the super-Yangian. Then, 
$v^+$ is also a highest weight vector for $T^{-1}(u)$:
\begin{eqnarray}
&& T_{kl}^{'(n)}\,v^+=0 \mb{for} k>l\,,\ 0<n \mb{i.e.} 
  T'_{kl}(u)\,v^+=0 \mb{for} k>l \label{t'-hw}\,,\\
&& T_{kk}^{'(n)}\,v^+=\lambda_{k}^{'(n)}\,v^+ \mb{for} 0<n \mb{i.e.} 
  T'_{kk}(u)\,v^+=\lambda'_{k}(u)\,v^+\,.  \label{t'-eigen}
\end{eqnarray}
\end{proposition}
\prf 
We make a recursion on $n$.
Applying (\ref{expr:t'}) for $n=1$ on $v^+$, it is easy to see that
(\ref{t'-hw}) and (\ref{t'-eigen}) are true for $n=1$. \\
Suppose now that we have for a given $s>0$ and
some scalars $\lambda_{k}^{'(n)}$
\begin{eqnarray}
&& T_{kl}^{'(n)}\,v^+=0 \mb{for} k>l\,,\ 0<n<s \nonu
&& T_{kk}^{'(n)}\,v^+=\lambda_{k}^{'(n)}\,v^+ \mb{for} 0<n<s\,,
\end{eqnarray}
Applying (\ref{expr:t'}) for $n=s$ and $k>l$ on $v^+$, one gets
\begin{eqnarray}
T'^{(s)}_{kl}\,v^+ &=& -\sum_{c=1}^{l} \sum_{p=1}^{s-1}
T'^{(s-p)}_{kc}T_{cl}^{(p)}\,v^+ 
                     = -\sum_{c=1}^{l} \sum_{p=1}^{s-1} \left[
T'^{(s-p)}_{kc}\,,T_{cl}^{(p)}  \right\}\,v^+= \nonumber\\
                   &=&
\sum_{a=1}^{l}(-1)^{[a]}\sum_{p=1}^{s-2}p \sum_{c=1}^{l}
\left[ T'^{(s-p-1)}_{kc}\,,T_{cl}^{(p)}  
\right\}\,v^+\,, \label{primo}
\end{eqnarray}
where to get the last equality, we have used (\ref{tt'-serie}).
Iterating $r$ times (with $2\leq r\leq s-1$)  this calculation
we are led to :
$$
T'^{(s)}_{kl}\,v^+  = A_{l,r} \sum_{p=1}^{s-r-1} B_{s,r,p}
\sum_{c=1}^{l} \left[ T'^{(s-p-r)}_{kc}\,,T_{cl}^{(p)}
\right\}\,v^+\,.
$$
where  $A_{l,r}$ and $B_{s,r,p}$ are some resummation numbers.
Taking $r=s-1$ gives (\ref{t'-hw}) for $n=s$, which is thus proven for
all $n$.\\
Finally, applying (\ref{expr:t'}) for $n=s$ and $k=l$ on $v^+$, we have:
\begin{eqnarray*}
T'^{(s)}_{kk}\,v^+ &=&
-\lambda^{(s)}_{k}\,v^+-\sum_{p=1}^{s-1}\lambda'^{(s-p)}_{k}\lambda_{k}^{(p)}\,v^+
+\\
   &&+ \sum_{c=1}^{k-1}(-1)^{[c]}
\sum_{p=1}^{s-1}p\left(\lambda'^{(s-p-1)}_{k}\lambda_{k}^{(p)}-
\lambda'^{(s-p-1)}_{c}\lambda_{c}^{(p)}\right) \,v^+ +\\
  &&+
\sum_{c=1}^{k-1}(-1)^{[c]}\sum_{p=1}^{s-2}\,p\left(\sum_{a=1}^{k-1}
  \left[T'^{(s-p-1)}_{ka}\,,T_{ak}^{(p)}\right\} - 
\sum_{a=1}^{k-1}\left[T_{ca}^{(p)}\,,T'^{(s-p-1)}_{ac}\right\}\right)\,v^+\,.
\end{eqnarray*}
Again, iterating as in eq. (\ref{primo}), we
see that only scalar terms
acting on $v^+$ will survive in the r.h.s.  This proves the property.
\finprf
It remains to determine the expression of the eigenvalues $\lambda'_k(u)$.
This is done in the following proposition:  
\begin{proposition}\label{prop:Tinv}
Let $\lambda'_k(u)$ be the
eigenvalue of $T^{-1}_{kk}(u)$ on $v^+$, $k=1,\dots,\cM+\cN$. We have
\begin{equation}\label{autovalori}
\lambda'_k(u) = \left\{\begin{array}{l}    
\frac{\lambda_1(u+\hbar)\cdots
\lambda_{k-1}(u+\hbar(k-1))}
{\lambda_1(u) \cdots
\lambda_{k}(u+\hbar(k-1)) }\,, \qquad k= 1,\dots,\cM\,,
\\[2.1ex]
Z(u) \frac{\lambda_{k+1}(u+\hbar(2\cM-k))\cdots\lambda_{\cM+\cN}(u+\hbar(\cM-\cN+1))}
{\lambda_k(u+\hbar(2\cM-k))\cdots\lambda_{\cM+\cN}(u+\hbar(\cM-\cN))}\,,
\qquad k=\cM+1,\ldots,\cM+\cN\,.
\end{array}\right.
\end{equation}
\end{proposition}
\prf
In order to find the first $\cM$ diagonal entries of $T^{-1}(u)$, we start writing
$$
\sum_{j \leq k} T_{ij}(u)T_{jk}^{-1}(u) \,v^+  = \delta_{ik}\,v^+\,,
$$
and taking $i,k \leq \cM$  we can write, in the distinguished grade, 
$$
\sum_{j \leq k} \left(T^{(\cM)}(u)\right)_{ij}
T_{jk}^{-1}(u) \,v^+  =
\delta_{ik}\,v^+\, \qquad i,k\leq \cM\,.
$$
Considering this as an identity in $\cY(\cM|\cN)[u^{-1}]\otimes
End(\CC^{\cM})$, we can act on the left with
$(T^{(\cM)}(u))^{-1}$, obtaining
\begin{equation}
 T_{kj}^{-1}(u) \,v^+  = \left(T^{(\cM)}(u)\right)_{kj}^{-1}
\,v^+\,, \qquad k,j = 1,...,\cM\,.
\label{eq:truc}
\end{equation}  
Let us stress that in (\ref{eq:truc}), $T_{kj}^{-1}(u)$ is the entry 
$(k,j)$ of the inverse of the $(\cM+\cN)\times(\cM+\cN)$ matrix $T(u)$, while  
$\left(T^{(\cM)}(u)\right)_{kj}^{-1}$ is the entry $(k,j)$ of the inverse of the
$\cM\times\cM$ matrix $T^{(\cM)}(u)$.
In particular, we get the relation
$$
T_{kk}^{-1}(u) \,v^+  = \lambda'^{(\cM)}_k(u)\,v^+\,, \qquad
k = 1,...,\cM\,
$$
where the $\lambda'^{(\cM)}_k(u)$ are the eigenvalues on $v^+$ of 
$\left(T^{(\cM)}(u)\right)_{kk}^{-1}$. 
It has been shown in \cite{molrag,byebye}
that these eigenvalues can be written as
\begin{equation}\label{foraut}
\lambda'^{(\cM)}_k(u) =
\frac{\lambda^{(\cM)}_{1}(u+\hbar)\cdots\lambda^{(\cM)}_{k-1}(u+\hbar(k-1))}
{\lambda^{(\cM)}_1(u)\cdots\lambda^{(\cM)}_{k}(u+\hbar(k-1))}\,,
\end{equation}
which leads to the first line of eq. (\ref{autovalori}).

For the last $\cN$ diagonal entries of $T^{-1}(u)$ we start 
writing in block form the relation $T^t(u)\left(T^{t}(u)\right)^{-1}\,v^+=v^+$, 
 setting
$$
T^t(u) = \left( \begin{array}{cc} 
\left(T^{(\cM)}(u)\right)^t & F(u)  \\
 G(u)  & \left(T^{(\cN)}_{-\hbar}(u)\right)^t
\end{array} \right)\,,\qquad
\left(T^{t}(u)\right)^{-1}\,v^+ =\left( \begin{array}{cc}  
A(u) & 0\\ *  & D(u)
\end{array} \right)\,v^+\,.
$$
We then  read from the lower right block
\begin{equation}\label{Auv}
D(u)\,v^+ = \left(\left(T^{(\cN)}_{-\hbar}(u)\right)^t\right)^{-1}\,v^+\,.
\end{equation}
The l.h.s. of this equation is computed via eq. (\ref{HAT}) which
implies, for $k > \cM$,
$$
\left( D(u)\right)_{k-\cM,k-\cM}\,v^+ =
\left(T^{t}(u)\right)^{-1}_{kk}\,v^+ =
\frac{1}{Z(u-\hbar(\cM-\cN))}\,T'_{kk}(u-\hbar(\cM-\cN))\,v^+\,. 
$$ 
The r.h.s. of the equation is computed via eq. (\ref{submatrix}). 
Comparing the left and right hand sides leads to
\begin{equation}\label{upperleft}
\lambda'_k(u) = z_{-\hbar}^{(\cN)}(u+\hbar(\cM-\cN))\,Z(u) \,\lambda'^{(\cN)}_{k-\cM}(u+\hbar
\cM)\,\qquad k=\cM+1,...,\cM+\cN\,,
\end{equation}
where the $\lambda'^{(\cN)}_k(u)$ are the eigenvalues on $v^+$ of 
diagonal elements
of the $T^{(\cN)}_{-\hbar}(u)$ matrix. 
Applying eq. (\ref{foraut}) to the $\cY_{-\hbar}(\cN)$
subalgebra, we can write these eigenvalues as 
$$
\lambda'^{(\cN)}_l(u) =
\frac{\lambda^{(\cN)}_{1}(u-\hbar)\cdots\lambda^{(\cN)}_{l-1}(u-\hbar(l-1))}
{\lambda^{(\cN)}_1(u)\cdots\lambda^{(\cN)}_{l}(u-\hbar(l-1))}\,,
\ l=1,...,\cN\,.\quad
$$
Inserting the value (\ref{eq:liouv-N}) of $z_{-\hbar}^{(\cN)}$
 in eq. (\ref{upperleft}) we find
the second line of eq. (\ref{autovalori}).
\finprf
In a finite dimensional irreducible representation, where relations (\ref{monic})
hold, we can rewrite eq. (\ref{autovalori}) in the following form:
$$\label{autovaloripol}
\lambda'_k(u) = \left\{\begin{array}{l}    
\frac{1}{\lambda_1(u)} \prod_{m=1}^{k-1} 
\frac{P_m(u+\hbar(m+1))}{P_m(u+\hbar m)}\,,\qquad k=1,\dots,\cM\,,
\\[2.1ex]
\frac{Z(u)}{\lambda_{\cM+\cN}(u+\hbar(\cM-\cN))} \prod_{m=k}^{\cM+\cN-1} 
\frac{P_m(u+\hbar(2\cM-m))}{P_m(u+\hbar(2\cM-m+1))}\,,\qquad k=\cM+1,\dots,\cM+\cN\,.
\end{array}\right.
$$

\subsubsection{Quantum Berezinian\label{qber}}
The quantum Berezinian was defined by Nazarov \cite{naza}. It plays a
similar role in the study of the Yangian  $\cY(\cM|\cN)$ 
as the quantum determinant does in the case of the Yangian $\cY(\cN)$.
\begin{definition}
The quantum Berezinian is the following power series with
coefficients in the Yangian  $\cY(\cM|\cN)$:
\begin{eqnarray}
Ber(u) &=& \sum_{\sigma \in S_{\cM}} \mathrm{sgn}(\sigma)\,
T_{\sigma(1)1}(u+\hbar(\cM-\cN-1))\cdots
T_{\sigma(\cM)\cM}(u-\hbar\cN)
\nonu
&&\label{ber}
\times  \sum_{\tau \in S_{\cN}} \mathrm{sgn}(\tau)\,
T^*_{\cM+\tau(1)\,,\cM+1}(u-\hbar\cN)\cdots
 T^*_{\cM +\tau(\cN)\,,\cM+\cN}(u-\hbar)\,.
\end{eqnarray}
\end{definition}
One can immediately recognize that
\begin{equation}\label{qdeTqdeT}
Ber(u) = \mathrm{qdet}\,T^{(\cM)}(u+\hbar(\cM-\cN-1))\,
\mathrm{qdet}\,T^{*(\cN)}(u-\hbar\cN)\,.
\end{equation}

\begin{proposition}\label{prop:central}\cite{naza}
The coefficients of the quantum Berezinian (\ref{ber}) are central in
$\cY(\cM|\cN)$.\\
They are related to the Liouville contraction through the identity
\begin{equation}\label{bZ}
Ber(u)\,Z(u) = Ber(u+\hbar)\,.
\end{equation}
\end{proposition}

The quantum Berezinian being central, one computes its value in the highest weight module
by
applying expression (\ref{ber}) to the h.w. vector $v^+$. We get
\begin{equation}\label{Berni}
Ber(u) = \prod_{l=1}^{\cM}
\lambda_l(u-\hbar \cN+\hbar(l-1))\,
\prod_{l=\cM+1}^{\cM+\cN}\lambda'_l(u-\hbar(\cM+\cN-l+1))\,,
\end{equation}
where the $\lambda'_l(u)$, $l=\cM+1,...,\cM+\cN$ are given in eq. (\ref{autovalori}). 
Substitution of this expression in the identity (\ref{bZ})
yields the following expression for $Z(u)$:
\begin{equation}\label{ZZ}
Z(u) = \frac{Ber(u+\hbar)}{Ber(u)}=
 \prod_{k=1}^{\cM}
\frac{\lambda_k(u+\hbar k)}{\lambda_k(u+\hbar(k-1))}
\prod_{l=\cM+1}^{\cM+\cN}\frac{\lambda_l(u+\hbar(2\cM-l))}{\lambda_l(u+\hbar(2\cM-l+1))}\,.
\end{equation}
Inserting now this expression into eq. (\ref{autovalori}), one obtains:
\begin{corollary}
The eigenvalues of the diagonal elements of $T^{-1}(u)$ on $v^+$ are given  by
\begin{equation}\label{autovalorimigliorati}
\lambda'_k(u) = \frac{\prod_{m=1}^{k-1} \lambda_m(u+\hbar c_m)   }
{\prod_{m=1}^{k} \lambda_m(u +\hbar c_{m-1})  }\,, \qquad k=1,\dots, \cM+\cN\,.
\end{equation}
where we set $c_m = \sum_{l=1}^{m}(-1)^{[l]}$, $m=1,\dots,\cM+\cN$, 
and $c_{0}=0$.
\end{corollary} 
Using expressions (\ref{Berni}) and (\ref{autovalorimigliorati}), 
one gets the value of the quantum Berezinian:
\begin{equation}\label{eq:central}
Ber(u) = \prod_{k=1}^{\cM}\lambda_k(u +\hbar(k-1))\,
\prod_{l=\cM+1}^{\cM+\cN} 
\frac{1}{\lambda_l(u+\hbar(2\cM-l))}\,.
\end{equation}

In what follows, we will also use a different expression for $Ber(u)$,
proved also in \cite{naza}:
\begin{eqnarray}
Ber^{-1}(u) &=& \sum_{\sigma \in S_{\cM}} \mathrm{sgn}(\sigma)\,
T^*_{\sigma(1)1}(u+\hbar(\cM-1))\cdots T^*_{\sigma(\cM)\cM}(u)\times
\label{b-1}  \\ 
& & \times\sum_{\tau \in S_{\cN}}  \mathrm{sgn}(\tau)\,
T_{\cM+\tau(1)\,,\cM+1}(u+\hbar(\cM-\cN))\cdots
T_{\cM+\tau(\cN)\,,\cM+\cN}(u+ \hbar(\cM-1))\,.
 \nonumber
\end{eqnarray}

Applying to both factors of expression (\ref{qdeTqdeT}) for the
quantum Berezinian the known identity (holding in $\cY_{\hbar}(\cN)$)
\begin{equation}\label{antis}
\mathrm{qdet}\,T(u)A_{\cN} = T_{\cN}(u-\hbar(\cN-1))\cdots
T_1(u)A_{\cN}\,,
\end{equation}
where $A_{\cN}$ is the normalized antisymmetrizer in the tensor space
$End(\mathbb C^{\cN})^{\otimes \cN}$,
we can write 
$$
Ber(u) A_{\cM}A_{\cN} = T_{\cM}^{(\cM)}(u-\hbar\cN)\cdot\cdot
T_1^{(\cM)}(u+\hbar(\cM-\cN-1))
     T_{\cM+\cN}^{*(\cN)}(u+\hbar')\cdot\cdot
T_{\cM+1}^{*(\cN)}(u+\hbar'\cN) A_{\cM}A_{\cN}\,,
$$
where we have set $\hbar' = -\hbar$ in the second quantum determinant. 
The $A_{\cM}$ and 
$A_{\cN}$ antisymmetrizers are both one--dimensional projectors
respectively acting on the 
tensor product of $\cM$ and $\cN$  copies of the auxiliary space, and
can
be written in terms of the $R$ matrices defining $\cY(\cM)$ and
$\cY_{\hbar'}(\cN)$:
\begin{eqnarray*}
&&\!\!A_{\cM} = \left( R_{12} \cdots R_{1\cM} \right) \cdots
R_{\cM-1,\cM}\,, \qquad R_{ij}=R^{(\cM)}_{ij}(u_i-u_j)\,,
\quad u_{i}-u_{i+1} = \hbar\,,
\\
&&\!\!A_{\cN} = \left( R'_{\cM+1,\cM+2} \cdot\cdot R'_{\cM+1,\cM+\cN} \right)
\cdot\cdot R'_{\cM+\cN-1,\cM+\cN}\,, 
\quad R'_{ij}=R^{(\cN),\hbar'}_{ij}(u'_i-u'_j)\,,\ 
u'_{i}-u'_{i+1} = \hbar'\,.\qquad
\end{eqnarray*}
Writing now $T^{(\cM)}(u) = \II^{(\cM)}T(u)\II^{(\cM)}$ and
$T^{*(\cN)}(u) = \II^{(\cN)}T^*(u)\II^{(\cN)}$, and 
setting $\Pi_{\cM|\cN}
=\left(\II^{\cM}   \right)^{\otimes \cM} \otimes 
\left(\II^{\cN}   \right)^{\otimes \cN}$, we get
\begin{eqnarray*}
Ber(u) A_{\cM}A_{\cN} &=&  \Pi_{\cM|\cN} T_{\cM}(u-\hbar\cN)\cdots
T_1(u+\hbar(\cM-\cN-1)) \times \\
  &&\times  T^*_{\cM+\cN}(u-\hbar)\cdots
T^*_{\cM+1}(u-\hbar\cN) A_{\cM}A_{\cN}\,.
\end{eqnarray*}
The same steps applied to eq. (\ref{b-1}) lead to the following
equation.
\begin{eqnarray*}
Ber^{-1}(u) A_{\cM}A_{\cN} &=&  \Pi_{\cM|\cN}
T'_{\cM}(u+\hbar(\cM-1))\cdots T'_1(u) \times \\
 &&\times  T_{\cM+\cN}(u+\hbar(\cM-1))\cdots
T_{\cM+1}(u+\hbar(\cM-\cN)) A_{\cM}A_{\cN}\,.
\end{eqnarray*}
The above expressions can be considered as the graded counterparts
of eq. (\ref{antis}): both relations act on a number
of copies of the auxiliary space equal to the dimension of the
Yangian and relate a $(\cM+\cN)$--fold tensor product of $T$
matrices to a central element by means of suitable one-dimensional
projectors.

\subsection{Reflection superalgebra}
To  study (soliton-preserving) open spin chains, we need to introduce another 
algebraic structure, the reflection algebra. It is a subalgebra of the 
super-Yangian, and actually can be defined from any quantum group. Focusing 
on the super-Yangian, the reflection superalgebra is a subalgebra of
$\cY(\cM|\cN)$, built as follows. One starts to consider 
\begin{equation}
B(u)=T(u)\,K(u)\,T^{-1}(-u)
\end{equation}
where $T(u)$ generates the super-Yangian and $K(u)$ is a matrix 
obeying the graded reflection (boundary Yang--Baxter) equation
\begin{equation}
  R_{12}(u_{1}-u_{2})\ K_{1}(u_{1})\ R_{21}(u_{1}+u_{2})\ K_{2}(u_{2})=
  K_{2}(u_{2})\ R_{12}(u_{1}+u_{2})\ K_{1}(u_{1})\ R_{21}(u_{1}-u_{2}) \;.
\label{reK}
\end{equation}
Using the exchange relations
(\ref{rtt}), it is easy to deduce that $B(u)$ also obeys the graded
reflection equation
\begin{equation}
  R_{12}(u_{1}-u_{2})\ B_{1}(u_{1})\ R_{21}(u_{1}+u_{2})\ B_{2}(u_{2})=
  B_{2}(u_{2})\ R_{12}(u_{1}+u_{2})\ B_{1}(u_{1})\ R_{21}(u_{1}-u_{2}) \;,
  \label{re}
\end{equation}
or, in components:
\begin{eqnarray}
\Big[ B_{ij}(u)\,, B_{kl}(v) \Big\} &=& \frac{(-1)^{\eta(i,j,k)}\hbar}{u-v}
\Big( B_{kj}(u)B_{il}(v)- B_{kj}(v)B_{il}(u) \Big)\nonumber \\
&+& \frac{\hbar}{u+v} \Big((-1)^{[j]}\delta_{jk} \sum_{a=1}^{\cM+\cN} B_{ia}(u)B_{al}(v)
- (-1)^{\eta(i,j,k)}\delta_{il}\sum_{a=1}^{\cM+\cN} B_{ka}(v)B_{aj}(u)
  \Big) \nonumber \\
&-& \frac{\hbar^2}{u^2-v^2}\delta_{ij} \Big( \sum_{a=1}^{\cM+\cN} B_{ka}(u)B_{al}(v)
- \sum_{a=1}^{\cM+\cN} B_{ka}(v)B_{al}(u)
  \Big)\,.\label{recomponents}
\end{eqnarray}
This relation shows that $B(u)$ generates a subalgebra 
of the super-Yangian, called reflection algebra and denoted $\fB$. 

Using the coproduct (\ref{coprod1}), 
one then shows that 
\begin{eqnarray}
\Delta\,\left(B_{ij}(u)\right) &=& 
\sum_{l,m=1}^{\cM+\cN} (-1)^{([m]+[j])([m]+[l])} T_{il}(u)T'_{mj}(-u) 
\otimes B_{lm}(u)\,.
\label{eq:coideal}
\end{eqnarray}
This proves that the reflection algebra is a Hopf coideal of
$\cY(\cM|\cN)$:
$$
\Delta \left( \fB   \right) \subseteq \cY(\cM|\cN)\otimes \fB\,.
$$
This will allow us to define monodromy matrices for open spin chains 
(see section \ref{monopen} below). In this context, 
the matrix $K(u)$ will be related to the boundary condition of the spin 
chain.
Hence, the classification of $K$ matrices is essential in the study of
open spin chains. As far as the super-Yangian is concerned, they have 
been classified in \cite{selene}. The result is summarized in the
following proposition
\begin{proposition}\label{reflectionclass}
Any invertible solution of the soliton preserving reflection equation
(\ref{reK}) takes the form
$K(u) = U\left(\EE   + \frac{\xi}{u}\II \right)U^{-1}$ where either
\begin{enumerate}
\item $\EE$ is diagonal and $\EE^2=\II$ (diagonalizable solutions)
\item $\EE$ is strictly triangular and $\EE^2 = 0$ (non--diagonalizable solutions) 
\end{enumerate}
The matrix $U$ is an element of the group $GL(\mathcal M) \times GL(\mathcal 
N)$, independent of the spectral parameter; 
 $\xi$ is a free parameter, and
 the classification is done up to multiplication by a function of the spectral parameter.
\end{proposition}
We will restrict to the case of diagonalizable solutions. The 
possible matrices $\EE$ are
 then labeled by two integers $L_{1}$ and $L_{2}$,  
$0\leq L_{1}\leq L_{2}\leq \cM+\cN$, which count the number 
of $-1$ on the diagonal of $\EE$:
$$
\EE = \mathrm{diag}\,(\underbrace{-1,\dots, -1}_{L_1}, 
\underbrace{1,\dots,1}_{L_2-L_1},\underbrace{-1,\dots, 
-1}_{\cN+\cM-L_{2}})\equiv
\mathrm{diag}\,(\theta_{1},\ldots,\theta_{\cM+\cN})\,.
$$
\textsl{Let us stress that
the diagonalization matrix $U$ being constant, it is sufficient 
 to consider diagonal $K(u)$ matrices: the other cases are recovered by a 
conjugation $T(u)\to U^{-1}\,T(u)\,U$ on each site of the chain, 
which does not affect the reflection algebra, nor the transfer matrix
\cite{selene}.}
The algebraic structure of $\fB$ does depend on the 
choice for $K(u)$. Indeed, from the expansion
\begin{equation}\label{expB}
B_{ij}(u) = \theta_{i}\,\delta_{ij} +\frac{1}{u}\,\left(
(\theta_{i}+\theta_{j})\,T^{(1)}_{ij}-\xi\,\delta_{ij}\right)+
\frac{1}{u^2}(\ldots)\,.
\end{equation}
we deduce that, when $L_{1}\leq \cM\leq L_{2}$,
 the Lie sub-superalgebra in $\fB$ is 
$gl(L_{1}|\cM+\cN-L_{2})\oplus gl(\cM-L_{1}|L_{2}-\cM)$.
 Hence, the notation $\fB$ should also contain the labels 
$\cN,\cM,L_{1},L_{2}$: we omit them for simplicity.

In the following, we will choose the normalisation of the resulting 
reflection matrix in such a way 
that its entries are analytical:
\begin{equation}\label{Kmat}
K(u) =  \mathrm{diag}\,(\underbrace{\xi-u,\dots, \xi-u}_{L_1\, \mathrm{terms}}, 
\underbrace{u+\xi,\dots,u+\xi}_{L_2-L_1 \mathrm{terms}},\xi-u,\dots, \xi-u)\,.
\end{equation}

\subsubsection{Highest weight representations of the reflection algebra}
We construct highest weight representations of
the reflection superalgebras based on those of the super-Yangian. This
construction will be used later on to build open spin chains. However,
a complete classification, similar to the
one done in \cite{molrag} for reflection algebras (based on the Yangian
$\cY(\cN)$), remains to be done. 

\begin{proposition}\label{Brepr}
The vector $v^+$ is a highest weight vector for the representations of the 
reflection algebra obtained from the
representation (\ref{triang}) of $\cY(\cM|\cN)$ with:
\begin{equation}
 B_{kl}(u)\,v^+ = 0\,,  \qquad \qquad 
1 \leq l < k \leq \mathcal M + \mathcal N\,, 
\label{eq:Bhw}
\end{equation}
\begin{equation}
 B_{kk}(u) \,v^+ = \frac{2u}{2u -\hbar c_{k-1}}\,g_k(u)
\,\lambda_k(u)\,\lambda'_k(-u)\,v^+ 
- \sum_{j=1}^{k-1}g_j(u)\,a_j(u)\,v^+\,,\ 
   1 \leq k \leq \mathcal M + \mathcal N\,,\quad
\end{equation}
where $c_k = \sum_{a=1}^{k}(-1)^{[a]}$ and
\begin{eqnarray}\label{gfunc}
g_k(u) &=& \left\{\begin{array}{ll} \xi -u \,, 
&   \mb{if} 1 \leq k \leq L_1 \\[1.2ex]
\xi + u -\hbar c_{L_1}\,,  &\mb{if} 
L_1 <  k \leq  L_2\,,\\[1.2ex]
 \xi-u -\hbar(c_{L_1} - c_{L_2})\,,  &   
\mb{if} L_2<k\leq \mathcal M + \mathcal N\,, 
 \end{array}\right.
\\[1.2ex]
a_k(u) &=& (-1)^{[k]}\,\hbar\, \frac{2u \la_k(u) 
\la'_k(-u)}{(2u-\hbar c_k)(2u - \hbar c_{k-1})}
\,.
\end{eqnarray}
\end{proposition}
\prf
We start writing, for $k > l$,
\begin{eqnarray}
B_{kl}(u)\,v^+ &=& \sum_{j = 1}^{l} 
T_{kj}(u) \,K_{jj}(u)\, T'_{jl}(-u)\,v^+ 
=\sum_{j=1}^{l}K_{jj}(u)\,
\left[ T_{kj}(u)\,, T'_{jl}(-u)  \right\}\,v^+\,.
\label{eigenb}
\end{eqnarray}
{From} the commutation relations, we find for $a\leq l<k$
\begin{equation}\label{intermed}
 \left[ T_{ka}(u)\,, T'_{al}(-u)  \right\}\,v^+\, = (-1)^{[a]}\frac{\hbar}{2u}
\sum_{b = 1}^{l}T_{kb}(u)\,T'_{bl}(-u)\,v^+\,,
\end{equation}
Considering the case $a=l$, we see that the l.h.s. of (\ref{intermed})
vanishes, so that
$$
\sum_{b = 1}^{l}T_{kb}(u)\,T'_{bl}(-u)\,v^+ = 0\,.
$$
Hence the right hand side of eq. (\ref{eigenb})  also vanishes,
proving (\ref{eq:Bhw}).\\
We now turn to the case $l=k$, i.e. to the eigenvalues of $B_{kk}(u)$ on $v^+$. 
We start defining 
$$
f_a(u) \doteq \sum_{k=1}^a T'_{ak}(-u)\,T_{ka}(u)\,v^+
\mb{and} \Psi_i(u) \doteq \sum_{k = 1}^i T_{ik}(u)\,T'_{ki}(-u)\,v^+\,.
$$
The supercommutation relations applied to these definitions imply 
\begin{equation}\label{systema} 
\left\{\begin{array}{l}
\displaystyle
 f_a(u) = \frac{1}{2u - \hbar\,c_{a-1}}\left(2u\lambda_a(u)\lambda'_a(-u)\,v^+- 
 \hbar\sum_{k=1}^{a-1}(-1)^{[k]}\Psi_k(u)\right)\\
 \displaystyle
 \Psi_a(u) = \frac{1}{2u -\hbar\,c_{a-1}}\left(2u\lambda_a(u)\lambda'_a(-u)\,v^+-
 \hbar\sum_{k=1}^{a-1}(-1)^{[k]}f_k(u)\right)\,, \end{array}\right.
\end{equation}
for $a=1,\dots,\mathcal M + \mathcal N$. Since $f_{1}(u) 
= \Psi_{1}(u) = \la_{1}(u)\la'_{1}(-u)\,v^+$, 
the system (\ref{systema}) has a unique solution $f_a(u) = \Psi_a(u)$, so  we can rewrite 
the expression of $f_a(u)$ as
\begin{equation}\label{ff}
\left( 1 -\frac{\hbar}{2u}\,c_{a-1} \right) f_a(u) 
= \la_a(u) \la'_a(-u)\,v^+ - \frac{\hbar}{2u}\sum_{k=1}^{a-1}(-1)^{[k]}f_k(u)\,.
\end{equation}
Eq.(\ref{ff}) is a triangular linear system in the unknowns $f_a(u)$ whose unique solution 
can be written as:
\begin{equation}\label{fff}
f_j(u) = \frac{\la_j(u)\la'_j(-u)}{1-\frac{\hbar}{2u}\,c_{j-1}}\,v^+ - 
\sum_{l=1}^{j-1}\frac{(-1)^{[l]}\hbar \la_{l}(u)\la'_{l}(-u)}{2u\left( 1-\frac{\hbar}{2u}
\,c_{l} \right)\left(1-\frac{\hbar}{2u}\,c_{l-1}\right)}\,v^+
= \frac{\la_j(u)\la'_j(-u)}{1-\frac{\hbar}{2u}\,c_{j-1}}\,v^+ - 
\sum_{l=1}^{j-1} a_{l}(u)\,v^+\,.
\end{equation}
Using this expression it is now clear that for $j\leq L_1$ we can write:
$$
B_{jj}(u)\,v^+ =  (\xi - u)f_j(u) 
= \left(\frac{2u(\xi -u)\la_j(u)\la'_j(-u)}{2u-\hbar c_{j-1}}-
(\xi-u)\sum_{k=1}^{j-1}a_k(u)\right)\,v^+\,.
$$
For $L_1 < j\leq L_2$ we have 
\begin{eqnarray}
B_{jj}(u)\,v^+ 
&=& (\xi+u) f_j(u) -2u \sum_{k=1}^{L_1} T_{jk}(u)T'_{kj}(-u)
\,v^+\nonu
&=& (\xi+u-\hbar c_{L_1}) f_j(u) + \hbar \sum_{k=1}^{L_1}(-1)^{[k]}
f_k(u)\,,\qquad\quad
\label{step}
\end{eqnarray}
where to get the last equality we have used supercommutation relations on $T_{jk}(u)T'_{kj}(-u)$.
Using now eq. (\ref{fff}), we get
$$
\hbar \sum_{k=1}^{L_1}(-1)^{[k]}f_k(u) = 
(2u- \hbar c_{L_1})\sum_{k=1}^{L_1}a_k(u)\,v^+\,.
$$
Substituting the above equation in eq. (\ref{step}), we get the required result. 
\\
An analogous calculation for the $j >  L_2$ case
leads to (\ref{gfunc}).\finprf

\section{Closed super-spin chains\label{sec:closed}}
\subsection{Monodromy and transfer matrices}\label{monodsec}
One defines the ($L$ sites) monodromy matrix $\mathcal T(u)$ as:
\begin{equation}
\mathcal T(u) = \Delta^{(L)}\left(T(u)\right) = T(u)\otimes 
T(u)\otimes \cdots \otimes T(u)\, \in \mathrm{End}(
\mathbb C^{(\cM|\cN)})\otimes
\left(\mathcal Y(\cM|\cN)   \right)^{\otimes l}\,.
\end{equation}
Applying an evaluation map on each term of this tensor product 
provides the `usual' monodromy matrix: the different sites correspond to  
the terms in the tensor product, and the evaluation map defines the 
`spin' (the representation) carried by the site.
Taking different representations of the super-Yangian allows to 
construct various type of closed super-spin chain models.

{From} the relation (\ref{rtt}), it is easy to show
that both the trace and the supertrace of the monodromy matrix 
\begin{equation}
t(u) = tr_{a}\,\cT(u)=\sum_{i=1}^{\cM+\cN} \mathcal
T_{ii}(u)\mb{and} st(u)=str_{a}\,\cT(u)=
\sum_{i=1}^{\cM+\cN} (-1)^{[i]} \mathcal
T_{ii}(u)
\end{equation}
 generate commutative families of operators:
\begin{equation}
\left[t(u) \,,t(v)\right] = 0\ \mb{and}
\left[st(u)\,,st(v)\right] =0\,.
\end{equation}
Note however that $t(u)$ and $st(u)$ do \und{not} commute one
with each other. Hence, they will generate different families
of commuting observables.

\subsection{Global invariance of transfer matrices}\label{invariance}

Taking the supertrace on the auxiliary space $1$ in relation
(\ref{rtt}), one is left with
\begin{equation}
\left[ X\,, st(u) \right] = 0\,,\qquad \forall\ X \in gl(\cM|\cN)\,.
\end{equation}
On the other hand, taking the trace in (\ref{rtt}) leads to
\begin{equation}
\left[ T_{kl}^{(1)}\,, t(u) \right] =
\left((-1)^{[i]}-(-1)^{[k]}\right)\,T_{kl}(u)\,,
\end{equation}
which is obviously zero iff $l$ and $k$ are both even or odd indices:
\begin{equation}
\left[ X\,, t(u) \right] = 0\,,\qquad \forall\ X \in gl(\cM)\oplus
gl(\cN)\,.
\end{equation}
Then, the transfer matrix $st(u)$ enjoys the full $gl(\cM|\cN)$ symmetry,
while the transfer matrix $t(u)$ is only $gl(\cM)\oplus gl(\cN)$
invariant. 

It is thus reasonable to think that the models associated to $st(u)$
are more relevant than the ones associated to $t(u)$ for the
construction of super-spin chain models. We will nevertheless present
the
Bethe anstaz for both transfer matrices. Note however that the
construction of open spin chain models is possible for the supertrace 
only, emphazising the difference between $t(u)$ and $st(u)$.

\subsection{Pseudovacuum for transfer matrices}
Starting from a $\cY(\cM|\cN)$ highest weight vector it is possible
to construct an eigenvector of the transfer matrix.
If $V_1,...,V_L$ are highest weight modules for $\mathcal Y(\cM|\cN)$,
with highest weight vectors $v_1,...,v_L$, then the vector
$v^+ \doteq v_1 \otimes ... \otimes v_L$
is a highest weight vector for the monodromy matrix, and thus an
eigenvector of the transfer matrices:
\begin{equation}
\mathcal T_{ij}(u)\,v^+ =
0\,,\quad 1\leq j<i\leq \cM+\cN\,,
\end{equation}
\begin{equation}\label{diag}
\mathcal T_{kk}(u)\,v^+ = \left(\prod_{n=1}^L\lambda_k^{[n]}(u)\right)
v^+ \doteq \lambda_k(u) v^+  \,. 
\end{equation}
Eq. (\ref{diag}) allows to compute the eigenvalue of $st(u)$:
\begin{eqnarray}
t(u)\,v^+ &=& \wh\Lambda_0(u)\,v^+\,,
\mb{with}
\wh\Lambda_0(u) \doteq \sum_{k=1}^{\cM+\cN}  \lambda_k(u) =
\sum_{k=1}^{\cM+\cN}  \prod_{n=1}^L\lambda_k^{[n]}(u)\,,\\
st(u)\,v^+ &=& \Lambda_0(u)\,v^+\,,
\mb{where}
\Lambda_0(u) \doteq \sum_{k=1}^{\cM+\cN}(-1)^{[k]}  \lambda_k(u) =
\sum_{k=1}^{\cM+\cN}(-1)^{[k]}  \prod_{n=1}^L\lambda_k^{[n]}(u)\,.
\qquad
\end{eqnarray}

Using evaluation representations (\ref{evaluation}),
 $ev_{\pi_{n}}$ for $ 1 \leq n \leq L$, with  highest weight 
$$
\lambda^{[i]}_{k}(u) = 1 + 
(-1)^{[k]}\frac{\hbar}{u-a_i}\mu^{[i]}_k\,,
$$
 we easily get the highest weight of the representation:
\begin{eqnarray*}
ev_{\vec{\pi}} \left(\cT_{kk}(u)\right)\,v^+ &=& \prod_{n=1}^L 
\left(  1 + (-1)^{[k]}\frac{\hbar}{u-a_n}\mu^{[n]}_k \right)\,v^+\,,\qquad
k=1,\dots,\cM+\cN\,, \qquad\\
ev_{\vec{\pi}} \left(st(u)\right) \,v^+ &=&
\sum_{k=1}^{\cM+\cN}(-1)^{[k]}\prod_{n=1}^L 
\left(  1 + (-1)^{[k]}\frac{\hbar}{u-a_n}\mu^{[n]}_k \right)\,v^+\,.
\end{eqnarray*}
It is important for what follows to remark that the above relations
imply that the entries
of the matrix $(u-a_{n})T(u)$ in a $ev_{\vec{\pi}}$ representation are
analytical.
{From} now on, we will  use for the local and monodromy matrices 
the normalizations:
\begin{equation}\label{normaliz}
T_k^{[n]}(u) \mapsto (u-a_n)\,T_k^{[n]}(u)\,, \qquad \mathrm{and} \quad
\cT(u) \mapsto \prod_{n=1}^L(u-a_n)\cT(u)\,,
\end{equation}
that ensure analyticity of their entries.
The transfer matrix will be accordingly normalized. 
Notice that with the normalization (\ref{normaliz}) the highest weight in the  
$ev_{\pi_{n}}$ representation reads:
\begin{equation}\label{lambdarin}
\lambda_k^{[n]}(u) = u -a_n + (-1)^{[k]}\,\hbar\,\mu_k^{[n]}
\mb{and} \lambda_k(u) =\prod_{n=1}^L 
\left(u -a_n + (-1)^{[k]}\,\hbar\,\mu_k^{[n]}\right) \,.
\end{equation}
Nevertheless, let us
stress the fact that the above calculation only relies  on the
existence 
of a highest weight vector, and thus remains valid for infinite
dimensional (highest weight) representations.
When the representations are finite dimensional, it is possible to
rewrite $\Lambda_0(u)$ in terms of Drinfeld polynomials. Indeed, we
will see that the BAEs depend on the representation only through the
Drinfeld polynomials.

\subsection{Dressing hypothesis}
Having determined the form of the pseudovacuum
eigenvalue we assume now the following form for the general transfer
matrix eigenvalues:
\begin{eqnarray}
\wh\Lambda(u) &=& \sum_{k=1}^{\cM+\cN}  \lambda_k(u)\,\wh
A_{k-1}(u)\,, \label{dressf0} \\
\Lambda(u) &=& \sum_{k=1}^{\cM+\cN}(-1)^{[k]}
\lambda_k(u)\,A_{k-1}(u)\,,  \label{dressf}
\end{eqnarray}
where the so-called dressing functions $A_i(u)$ and $\wh A_{i}(u)$, 
$i=0,...,\cM+\cN-1$ are to be determined 
implementing a number of constraints upon the spectrum:
\begin{enumerate}
 \item the $R$ matrix and monodromy matrix being written in terms of
rational functions of the spectral parameter $u$, one assumes that 
$A_{l}(u)$, $\forall\,l$, are also rational functions; 
 \item analyticity requirements imposed on the spectrum lead to  the
assumption that $A_l(u)$ (resp. $\wh A_{l}(u)$)
has common poles with $A_{l\pm1}(u)$ (resp. $\wh A_{l\pm1}(u)$) only;
 \item the poles of the dressing functions will be
assumed simple: the relation between $A_l(u)$ and $A_{l+1}(u)$ poles 
is the simplest one which ensures the analyticity of the eigenvalues;
 \item the asymptotic expansion of the transfer matrix will provide
information about the number of factors in the aforementioned rational
functions;
 \item the generalized fusion provides  relations among the dressing
functions.
\end{enumerate}  
Requirements $1.$ and $2.$ fix the following form for the dressing
functions:
\begin{equation}\label{hypdress}
A_l(u) =  \left\{\begin{array}{ll} 
\displaystyle \prod_{j=1}^{M^{(l)}}
\frac{u-\alpha_j^{(l)}}{u-u_j^{(l)}-\hbar\frac{l}{2}}
         \prod_{j=1}^{M^{(l+1)}}
\frac{u-\beta_j^{(l+1)}}{u-u_j^{(l+1)}-\hbar\frac{l+1}{2}}\,,
         & 0\leq l<\cM\,, \\[1.2ex]
\displaystyle\prod_{j=1}^{M^{(l)}}
\frac{u-\alpha_j^{(l)}}{u-u_j^{(l)}-\hbar\,\left(\cM-\frac{l}{2}\right)}
       \prod_{j=1}^{M^{(l+1)}}
\frac{u-\beta_j^{(l+1)}}{u-u_j^{(l+1)}-\hbar\,\left(\cM-\frac{l+1}{2}\right)}\,,
       & \cM\leq l<\cM+\cN\,,   
\end{array}\right.
\end{equation}
where $M^{(0)}=M^{(\cM+\cN)}=0$, 
while the values of the integers $M^{(l)}$, $l= 1,\dots,\cM+\cN-1$
are to be
determined by means of 
asymptotic expansion (point $4.$ above), as will be shown in the next
section; 
the shifts in the denominators have been introduced for later
convenience.

The next step consists in finding constraints to determine
$\alpha_j^{(l)}$ and $\beta_j^{(l)}$
in terms of $u_j^{(l)}$. This is achieved by means of the generalized
fusion procedure. 

\subsubsection{Values of the $gl(\cM|\cN)$ Cartan generators}

As we have seen in section \ref{invariance}, the generators of the
finite--dimensional 
$gl(\cM|\cN)$ superalgebra commute with the transfer matrix. It is
thus possible to relate
the integers $M^{(l)}$, $l=1,\dots,\cM+\cN-1$, appearing
in the $\Lambda(u)$ dressing to the eigenvalues of the Cartan
generators of $gl(\cM|\cN)$.
This can be done in the following way.

Taking first the $u \rightarrow \infty$ in the expression
$(\ref{dressf})$ for $\Lambda(u)$
for an $L$ sites chain, one gets
$$
\Lambda(u) \sim u^L(\cM -\cN) +
u^{L-1}\sum_{k=1}^{\cM+\cN}(-1)^{[k]}\hbar\left( \lambda^{(1)}_k -
M^{(k-1)} + M^{(k)}  \right)\,,
$$
where we set $\lambda_k(u) = u + \hbar\lambda^{(1)}_k + O\left(
\frac{1}{u} \right)$.
On the other hand, the same expansion performed on the transfer
matrix $st(u)$  leads to 
$$
st(u) \sim u^L(\cM - \cN)+ u^{L-1}\sum_{k=1}^{\cM+\cN}\hbar\left(
\sum_{n=1}^{L} \cE_{k}^{[n]}\right)\,,
$$
where $\sum_{n=1}^{L} \cE_{k}^{[n]} = \sum_{n=1}^{L}(-1)^{[k]}
T^{(1)[n]}_{kk}$ is the $k$-th diagonal
generator of the global $gl(\cM|\cN)$ symmetry algebra of the chain.
Starting then from a transfer matrix eigenvector with eigenvalue 
(\ref{dressf}), one can write 
$$
(-1)^{[k]}h_k = \lambda^{(1)}_k - M^{(k-1)} + M^{(k)}\,,
$$
where $h_k$ is the eigenvalue of the diagonal generator
$\sum_{n=1}^{L} \cE_{k}^{[n]} $. 
For the Cartan generators of $gl(\cM|\cN)$, $s_k = (-1)^{[k]}\mathcal
E_k - (-1)^{[k+1]}\mathcal E_{k+1}$,
one gets
$$
s_k\,v = \left( 2 M^{(k)} - M^{(k-1)}-  M^{(k+1)}+ \lambda^{(1)}_k -
\lambda^{(1)}_{k+1}  \right)\,v\,.
$$
The above calculation shows that
the  values of the $M^{(k)}$ integers are
related to the conserved charges of the global symmetry of the chain: 
one must then take care that simplifications in the dressing functions
resulting from the fusion procedure
do not change their number of factors.
In other words each $M^{(k)}$ should
be kept independent from each other and only relations between the other parameters 
appearing in the dressing
are allowed, as we will shown in the next section.

\subsubsection{Generalized fusion from quantum Berezinian}

The relations (\ref{rt*t*}), (\ref{rt*t}) and (\ref{rtt*}), between
$T^*(u)$ and $T(v)$ show that we can define
another transfer matrix $st^*(u)=str T^*(u)$ which obeys
\begin{equation}
{[st(u)\,,\,st^*(v)]}=0 \mb{and} {[st^*(u)\,,\,st^*(v)]}=0
 \end{equation}
so that one can consider the dressing of $st^*(u)$ simultaneously
with 
the one of $st(v)$:
\begin{equation}\label{dressstar}
\Lambda^*(u) =
\sum_{k=1}^{\cM+\cN}(-1)^{[k]}\lambda_k^*(u)A^*_k(u)\,,
\end{equation}
where $ T^*_{kk}(u)\,v^+ = \lambda_k^*(u)\,v^+$
and
$$
A^*_l(u) =  \left\{\begin{array}{ll} 
\displaystyle\prod_{j=1}^{M^{(l)}}
\frac{u-\alpha_j^{*(l)}}{u-u_j^{*(l)}-\hbar\,\left(\cM-\frac{l}{2}\right)}
       \prod_{j=1}^{M^{(l+1)}}
\frac{u-\beta_j^{*(l+1)}}{u-u_j^{*(l+1)}-\hbar\,\left(\cM-\frac{l+1}{2}\right)}\,,
       \quad& 0\leq l<\cM\,,   \\[2.1em]
\displaystyle\prod_{j=1}^{M^{(l)}}
\frac{u-\alpha_j^{*(l)}}{u-u_j^{*(l)}-\hbar\frac{l}{2}}
	 \prod_{j=1}^{M^{(l+1)}}
\frac{u-\beta_j^{*(l+1)}}{u-u_j^{*(l+1)}-\hbar\frac{l+1}{2}}\,,
	 & \cM\leq l<\cM+\cN\,. 
\end{array}\right.
$$

Let $A_{\cM}$, $A_{\cN}$, $\Pi_{\cM|\cN}$ be the one--dimensional
projectors defined in section \ref{qber}
which act on auxiliary spaces $1,\dots,\cM+\cN$ and denote   
$$
\cT \cT^* = \cT_{\cM}(u-\hbar\cN)\cdots \cT_1(u+\hbar(\cM-\cN-1))
\cT^*_{\cM+\cN}(u-\hbar)\cdots \cT^*_{\cM+1}(u-\hbar\cN)\,. 
$$
 Then, from the following relation
\begin{equation}\label{fusB}
\cT\cT^* =  
Ber(u)A_{\cM}A_{\cN} + (1-\Pi_{\cM|\cN}) \cT\cT^*A_{\cM}A_{\cN} + 
    \cT \cT^*(1-A_{\cM}A_{\cN})\,,
\end{equation}
we deduce, by taking the supertrace in the spaces $1,\dots,\cM+\cN$,
that
$$
st(u-\hbar\cN)\cdots st(u+\hbar(\cM-\cN-1)) st^*(u-\hbar)\cdots
st^*(u-\hbar\cN) = (-1)^{\cN}Ber(u) + st_{\mathfrak f}^{(1)}(u)\,,
$$ 
where $st_{\mathfrak f}^{(1)}(u) = str_{1...\cM+\cN} \left[
(1-\Pi_{\cM|\cN}) \cT\cT^*A_{\cM}A_{\cN} + 
    \cT \cT^*(1-A_{\cM}A_{\cN})  \right]$ is a so--called fused
transfer matrix.
Then, acting with relation (\ref{fusB}) on any ($st(u)$ and $st^*(u)$)
eigenvector $v$ with eigenvalues $\Lambda(u)$, $\Lambda^*(u)$, 
one obtains
\begin{eqnarray}
&& \Lambda(u-\hbar\cN)\cdots \Lambda(u+\hbar(\cM-\cN-1))
\Lambda^*(u-\hbar)\cdots \Lambda^*(u-\hbar\cN) =
\nonumber \\
&=& (-1)^{\cN}\prod_{k=1}^{\cM}\lambda_k(u -\hbar(\cN-k+1))
\prod_{l=\cM+1}^{\cM+\cN}\lambda'_l(u+\hbar(\cM+\cN-l+1))+
\Lambda_{\mathfrak f}^{(1)}(u)\,, \label{fus2}
\end{eqnarray}
where $\Lambda_{\mathfrak f}^{(1)}(u)\,v = st_{\mathfrak
f}^{(1)}(u)\,v $ and we have used eq. (\ref{Berni}). Let us remark
that this relation shows that $v$ is also an eigenvector of
$t_{\mathfrak f}^{(1)}(u)$.
Using the postulated expression (\ref{dressf}) for the eigenvalues
and picking the term proportional to 
$\prod_{k=1}^{\cM}\lambda_k(u -\hbar(\cN-k+1))
\prod_{l=\cM+1}^{\cM+\cN}\lambda'_l(u+\hbar(\cM+\cN-l+1))$
in eq. (\ref{fus2}), we deduce a first constraint between the dressing
functions, namely
\begin{equation}\label{fus3}
A_0(u-\hbar\cN)\cdots
A_{\cM-1}(u+\hbar(\cM-\cN-1))A_{\cM}^*(u-\hbar\cN)\cdots A_{\cM+\cN-1}^*(u-\hbar) = 1\,.
\end{equation}
The simplest non--trivial choice of the
$\alpha_j^{(k)},\alpha_j^{*(k)}$ and $\beta_j^{(k)},\beta_j^{*(k)}$
satisfying this constraint is to
set $\alpha_j^{(k)} = u_j^{(k)} +\frac{\hbar}{2}(k+2) $,
$\beta_j^{(k+1)} =  u_j^{(k+1)} +\frac{\hbar}{2}(k-1)$, $\forall\,j$,
for $k=0,\dots ,\cM-1$, $u_j^{*(\cM)}=u_j^{(\cM)}-\hbar \cM$, and
$\alpha_j^{*(k)}
= u_j^{*(k)} +\frac{\hbar}{2}(k+2) $,  $\beta_j^{*(k+1)} =
u_j^{*(k+1)} +\frac{\hbar}{2}(k-1)$, $\forall\,j$,
for $k=\cM,\dots \cM+\cN-1$ 
in such a way that 
\begin{eqnarray*}
A_k(u) &=& \prod_{j=1}^{M^{(k)}}
\frac{u-u_j^{(k)}-\hbar\frac{k+2}{2}}{u-u_j^{(k)}-\hbar\frac{k}{2}}
    \prod_{j=1}^{M^{(k+1)}}
\frac{u-u_j^{(k+1)}-\hbar\frac{k-1}{2}}{u-u_j^{(k+1)}-\hbar\frac{k+1}{2}}\,,
\qquad\quad k=0,\dots,\cM-1\,, 
\\
A^*_k(u) &=& \prod_{j=1}^{M^{(k)}}
\frac{u-u_j^{*(k)}-\hbar\frac{k+2}{2}}{u-u_j^{*(k)}-\hbar\frac{k}{2}}
    \prod_{j=1}^{M^{(k+1)}}
\frac{u-u_j^{*(k+1)}-\hbar\frac{k-1}{2}}{u-u_j^{*(k+1)}-\hbar\frac{k+1}{2}}\,,
\qquad k=\cM,\dots,\cM+\cN-1\,, 
\end{eqnarray*}
and cancelations occur between  dressing functions labeled by
consecutive indices in expression (\ref{fus3}).
To fix the values of the $\alpha_j^{(k)}$ and $\beta_j^{(k)}$ for
$k\geq \cM$ we start setting
$$
\cT'\cT = \cT'_{\cM}(u+\hbar(\cM-1))\cdots \cT'_1(u)
\cT_{\cM+\cN}(u+\hbar(\cM-1))\cdots \cT_{\cM+1}(u+\hbar(\cM-\cN)) 
$$
and supertracing in all auxiliary spaces the identity
\begin{equation}\label{fus5}
\cT'\cT =  
Ber^{-1}(u)A_{\cM}A_{\cN} + (1-\Pi_{\cM|\cN}) \cT'\cT A_{\cM}A_{\cN} + 
    \cT' \cT(1-A_{\cM}A_{\cN})\,,
\end{equation}
we get
$$
st^*(u+\hbar(\cM-1))\cdots st^*(u) st(u+\hbar(\cM-1))\cdots
st(u+\hbar(\cM-\cN)) = (-1)^{\cN} Ber^{-1}(u) + st_{\mathfrak
f}^{(2)}(u)\,,
$$
where $st_{\mathfrak f}^{(2)}(u) = str_{1...\cM+\cN} \left[
(1-\Pi_{\cM|\cN}) \cT'\cT A_{\cM}A_{\cN} + 
    \cT' \cT(1-A_{\cM}A_{\cN})  \right] $. 
Acting again with the above equation on $v$, one obtains
\begin{eqnarray}
&& \Lambda^*(u+\hbar(\cM-1))\cdots \Lambda^*(u)
\Lambda(u+\hbar(\cM-1))\cdots \Lambda(u+\hbar(\cM-\cN)) =
\nonumber \\
&=& (-1)^{\cN}\prod_{l=1}^{\cM}
\lambda'_l(u+\hbar(\cM-l))\,\prod_{l=\cM+1}^{\cM+\cN}\lambda_l(u+\hbar(2\cM-l))+
\Lambda_{\mathfrak f}^{(2)}(u)\,, \label{fus6}
\end{eqnarray}
where $\Lambda_{\mathfrak f}^{(2)}(u)\,v = t_{\mathfrak
f}^{(2)}(u)\,v $ and  eq. (\ref{eq:central}) has been used. 
Picking up the term proportional to 
$\lambda'_l(u+\hbar(\cM-l))\,\prod_{l=\cM+1}^{\cM+\cN}\lambda_l(u+\hbar(2\cM-l))$,
we get
a second constraint on the dressing functions:
\begin{equation}\label{fus7}
A^*_0(u+\hbar(\cM-1))\cdots
A^*_{\cM-1}(u)A_{\cM}(u+\hbar(\cM-1))\cdots
A_{\cM+\cN-1}(u+\hbar(\cM-\cN))=1\,.
\end{equation}
To satisfy this second constraint we set
$ \alpha_j^{(k)}= u_j^{(k)} +\hbar(\cM-\frac{k}{2}-1) $, $
\beta_j^{(k+1)} = u_j^{(k+1)} +\hbar(\cM-\frac{k-1}{2})$
for $k = \cM,\dots,\cM+\cN-1$, and
$ \alpha_j^{*(k)}= u_j^{*(k)} +\hbar(\cM-\frac{k}{2}-1)$, $
\beta_j^{*(k+1)} = u_j^{*(k+1)} +\hbar(\cM-\frac{k-1}{2})$
for $k = 0,\dots,\cM-1$, so that
$$
A_k(u) = \prod_{j=1}^{M^{(k)}} \frac{u-u_j^{(k)}-\hbar\left(\cM
-\frac{k}{2} -1  \right)}
{u-u_j^{(k)}-\hbar\left(\cM -\frac{k}{2}\right)}
    \prod_{j=1}^{M^{(k+1)}}
\frac{u-u_j^{(k+1)}-\hbar\left(\cM-\frac{k-1}{2}\right)}
    {u-u_j^{(k+1)}-\hbar\left(\cM-\frac{k+1}{2}\right)}\,,
\quad \cM\leq k<\cM+\cN\,, 
$$
$$
A^*_k(u) = \prod_{j=1}^{M^{(k)}} \frac{u-u_j^{*(k)}-\hbar\left(\cM
-\frac{k}{2} -1  \right)}
{u-u_j^{*(k)}-\hbar\left(\cM -\frac{k}{2}\right)}
     \prod_{j=1}^{M^{(k+1)}}
\frac{u-u_j^{*(k+1)}-\hbar\left(\cM-\frac{k-1}{2}\right)}
    {u-u_j^{*(k+1)}-\hbar\left(\cM-\frac{k+1}{2}\right)}\,,
\qquad  0\leq k<\cM\,.     
$$
Again, it is seen that $u_j^{*(\cM)} = u_j^{(\cM)}-\hbar\cM$.

\begin{rmk}
Relations (\ref{fus3}) and (\ref{fus7}) also hold when the $A_l(u)$,
$A^*_l(u)$ functions are
replaced with $\wh A_l(u)$, $\wh A^*_l(u)$, thus leading to the same
form for the
dressing functions appearing in the  eigenvalues (\ref{dressf0}) and
(\ref{dressf}). 
\end{rmk}
\begin{rmk}
Using the $c_k$ integers introduced in proposition \ref{Brepr},
 one can write a single expression for the dressing functions:
$$
A_k(u) = \prod_{j=1}^{M^{(k)}} \frac{u-u_j^{(k)}-\frac{\hbar}{2}
\left(c_{k+1}+(-1)^{[k+1]}\right)}{u-u_j^{(k)}-\frac{\hbar}{2}c_k}
\ \prod_{j=1}^{M^{(k+1)}}
\frac{u-u_j^{(k+1)}-\frac{\hbar}{2}\left(c_k-(-1)^{[k+1]}\right)}
    {u-u_j^{(k+1)}-\frac{\hbar}{2}c_{k+1}}\,,
$$
$$
A^*_k(u) = \prod_{j=1}^{M^{(k)}} \frac{u-u^{*(k)}_j-\frac{\hbar}{2}(2\cM-c_{k+1}-(-1)^{[k+1]})}
                                    {u-u^{*(k)}_j-\frac{\hbar}{2}(2\cM-c_{k})}\,
           \prod_{j=1}^{M^{(k)}} \frac{u-u^{*(k+1)}_j-\frac{\hbar}{2}(2\cM-c_{k-1})}
                                    {u-u^{*(k+1)}_j-\frac{\hbar}{2}(2\cM-c_{k+1})}\,,
$$
\begin{eqnarray}
k &=& 1, \dots, \cM+\cN-1\,, \label{eq:dress-ck}\,.
\end{eqnarray}
\end{rmk}

\subsection{Bethe equations of closed spin chains}\label{BE1}
We have seen in the previous section that $A_{l}(u)=\wh A_{l}(u)$,
and that they have  the form
\begin{eqnarray*}
A_l(u) &=& \prod_{k=1}^{M^{(l)}}
\frac{u-u_k^{(l)}-\hbar\,\frac{l+2}{2}}{u-u_k^{(l)}-\hbar\,\frac{l}{2}}
        \prod_{k=1}^{M^{(l+1)}}
\frac{u-u_k^{(l+1)}-\hbar\,\frac{l-1}{2}}{u-u_k^{(l+1)}-\hbar\,\frac{l+1}{2}}\,,
       \qquad  0\leq l<\cM\,, \\
A_l(u) &=& \prod_{k=1}^{M^{(l)}}
\frac{u-u_k^{(l)}-\hbar\,\left(\cM-\frac{l}{2}-1\right)}{u-u_k^{(l)}-\hbar\,
\left(\cM-\frac{l}{2}\right)}
       \prod_{k=1}^{M^{(l+1)}}
\frac{u-u_k^{(l+1)}-\hbar\,\left(\cM-\frac{l-1}{2}\right)}{u-u_k^{(l+1)}-\hbar\,
\left(\cM-\frac{l+1}{2}\right)}\,,
       \quad \cM\leq l<\cM+\cN\,,   
\end{eqnarray*}
with the convention $M^{(0)}=M^{(\cM+\cN)}=0$. 

In order to establish analyticity of all eigenvalues of $\Lambda(u)$
and of $\wh\Lambda(u)$,
one imposes that the  residues  of $\Lambda(u)$ and $\wh\Lambda(u)$
at $u=u_j^{(n)}+\hbar\,\frac{n}{2}$ for $1\leq j\leq M^{(n)}$,
$0<n<\cM$,
and at $u=u_j^{(n)}+\hbar\,(\cM-\frac{n}{2})$ for $1\leq j\leq
M^{(n)}$, 
$\cM\leq n\leq \cM+\cN-1$, all vanish.

Introducing the function
\begin{equation}\label{epsilon}
\fe_n(u) \doteq
\frac{u-\hbar\,\frac{n}{2}}{u+\hbar\,\frac{n}{2}}, 
\end{equation}
the vanishing of these
residues leads to the following (Bethe ansatz) equations:
\begin{eqnarray}
&& \prod_{k=1}^{M^{(n-1)}}\fe_{-1}(u^{(n)}_j-u^{(n-1)}_k) 
\prod_{k \neq j}^{M^{(n)}} \fe_{2}(u^{(n)}_j-u^{(n)}_k)
\prod_{k=1}^{M^{(n+1)}}\fe_{-1}(u^{(n)}_j-u^{(n+1)}_k) =
\frac{\lambda_{n}(u_j^{(n)}+\hbar\frac{n}{2})}
{\lambda_{n+1}(u_j^{(n)}+\hbar\frac{n}{2})}\,, \nonu
&& \qquad 1\leq j\leq M^{(n)}\ ,\ 0<n<\cM \\[1.2ex]
&& \prod_{k=1}^{M^{(n-1)}}\fe_{1}(u^{(n)}_j-u^{(n-1)}_k) 
\prod_{k \neq j}^{M^{(n)}} \fe_{-2}(u^{(n)}_j-u^{(n)}_k)
\prod_{k=1}^{M^{(n+1)}}\fe_{1}(u^{(n)}_j-u^{(n+1)}_k) =
\frac{\lambda_{n}(u_j^{(n)}+\hbar(\cM-\frac{n}{2}))}
{\lambda_{n+1}(u_j^{(n)}+\hbar(\cM-\frac{n}{2}))}\,, \nonu
&& \qquad 1\leq j\leq M^{(n)}\ ,\ \cM<n<\cM+\cN \\[1.2ex]
&&
\prod_{k=1}^{M^{(\cM-1)}}\fe_{-1}(u^{(\cM)}_j-u^{(\cM-1)}_k) 
\prod_{k=1}^{M^{(\cM+1)}}\fe_{1}(u^{(\cM)}_j-u^{(\cM+1)}_k) =
\pm\ \frac{\lambda_{\cM}(u_j^{(\cM)}+\hbar\frac{\cM}{2})}
{\lambda_{\cM+1}(u_j^{(\cM)}+\hbar\frac{\cM}{2})}\,, \nonu
&& \qquad 1\leq j\leq M^{(\cM)}
\end{eqnarray}
where in the last equation the $+$ sign (resp. $-$ sign) corresponds
to the $\Lambda(u)$ BAE (resp. $\wh\Lambda(u)$ BAE). One recognizes in
the left-hand side of the BAEs the Cartan matrix of the $gl(\cM|\cN)$
superalgebra, while the right-hand side reflects the super-Yangian 
representation(s) spanned by the spin chain. 

When the representations are finite dimensional, the right-hand side
of these equations can be re-expressed in terms of Drinfeld
polynomials.
For instance, for the first set of BAEs, one gets 
\begin{equation}
\frac{\lambda_{n}(u_j^{(n)}+\hbar\frac{n}{2})}
{\lambda_{n+1}(u_j^{(n)}+\hbar\frac{n}{2})}
=\frac{P_{n}(u_j^{(n)}+\hbar\frac{n+2}{2})}
{P_{n}(u_j^{(n)}+\hbar\frac{n}{2})}
\mb{where}
    P_{i}(u) = \prod_{n=1}^L  P^{[n]}_{i}(u) \,,
\end{equation}
$P^{[n]}_{i}(u) $ being the Drinfeld polynomials for each site.

\section{Bethe equations for arbitrary Dynkin diagrams\label{sec:arbitraryDD}}
As already mentioned, up-to-now we have worked with the distinguished 
Dynkin diagram and its associated gradation. However, several Dynkin 
diagrams can be used to describe the same superalgebra, leading to 
inequivalent Dynkin diagram, and thus to different presentations of 
the Bethe equations. For each of the grading (i.e. for each 
inequivalent Dynkin diagram), 
one can apply the above procedure to 
determine the form of the dressing functions. This has been 
noticed in \cite{selene} for 
open super-spin chains in the fundamental representation of 
$sl(\cM|\cN)$. We generalize it for arbitrary super-spin chains.
The dressing functions keep essentially 
the same structure, with the following rules.

The inequivalent Dynkin diagrams of the $sl(\cM|\cN)$ superalgebras
contain only bosonic roots of same square length ("white dots"), 
normalized to 2, and isotropic fermionic roots ("grey dots"), which 
square to zero. A given
diagram is completely characterized by the $p$-uple of integers
$0<n_{1}<\ldots<n_{p}<\cM+\cN$ labelling the positions of the grey
dots of the diagram:
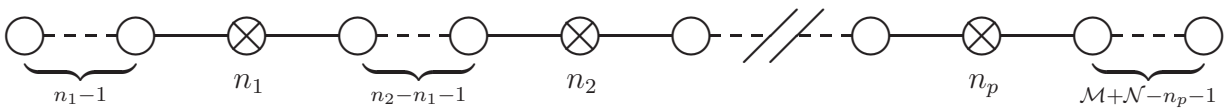
\begin{center}
\begin{picture}(452,20) \thicklines
   \multiput(0,10)(42,0){7}{\circle{14}}
   \multiput(320,10)(42,0){4}{\circle{14}}
   \put(79,5){\line(1,1){10}}\put(79,15){\line(1,-1){10}}
   \put(205,5){\line(1,1){10}}\put(205,15){\line(1,-1){10}}
   \put(357,5){\line(1,1){10}}\put(357,15){\line(1,-1){10}}
   \put(7,10){\line(1,0){4}}\put(15,10){\line(1,0){4}}
   \put(23,10){\line(1,0){4}}\put(31,10){\line(1,0){4}}
   \put(49,10){\line(1,0){28}}
   \put(91,10){\line(1,0){28}}
   \put(133,10){\line(1,0){4}}\put(141,10){\line(1,0){4}}
   \put(149,10){\line(1,0){4}}\put(157,10){\line(1,0){4}}
   \put(175,10){\line(1,0){28}}
   \put(217,10){\line(1,0){28}}
   \put(259,10){\line(1,0){4}}\put(267,10){\line(1,0){4}}
   \put(275,10){\line(1,0){4}}
   \put(272,0){\line(1,1){20}}\put(282,0){\line(1,1){20}}
   \put(293,10){\line(1,0){4}}
   \put(301,10){\line(1,0){4}}\put(309,10){\line(1,0){4}}
   \put(327,10){\line(1,0){28}}
   \put(369,10){\line(1,0){28}}
   \put(411,10){\line(1,0){4}}\put(419,10){\line(1,0){4}}
   \put(427,10){\line(1,0){4}}\put(435,10){\line(1,0){4}}
   \put(79,-10){$n_{1}$}
   \put(205,-10){$n_{2}$}
    \put(357,-10){$n_{p}$}
    \put(0,0){$\underbrace{\hspace{42pt}}_{n_{1}-1}$}
    \put(128,0){$\underbrace{\hspace{42pt}}_{n_{2}-n_{1}-1}$}
   \put(399,0){$\underbrace{\hspace{42pt}}_{\cM+\cN- n_{p}-1}$}
\end{picture}
\end{center}
\vspace*{\baselineskip}
where the total number of (grey and white) dots is $\cM+\cN-1$.
Formally, we define $n_0=0$ and $n_{p+1}=\cM+\cN$ although there is
actually no root at these positions.
Such a diagram defined by the $p$-uple $(n_i)_{i=1\dots p}$
corresponds to the superalgebra $sl(\cM|\cN)$ with
\begin{equation}
  \cM = \sum_{\atopn{i \textrm{ odd}}{ i\leq p+1}} n_i
  - \sum_{\atopn{i \textrm{ even}}{i< p+1}} n_i
  \qquad  \mbox{and}  \qquad
  \cN = \sum_{\atopn{i \textrm{ even}}{i\leq p+1}} n_i
  - \sum_{\atopn{i \textrm{ odd}}{i< p+1}} n_i \;.
\end{equation}
Accordingly, the $\ZZ_{2}$-grading is defined by
\begin{equation}
{[j]} = \frac{1-(-1)^{k}}{2}\,,\mb{i.e.}
(-1)^{[j]}=(-1)^{k}\,,
\mb{for} n_{k}+1\leq j\leq
n_{k+1}\,,\quad 0\leq k\leq\cM+\cN \,.
\end{equation}
For each of these gradings, one can compute a new value for the 
parameters
\begin{equation}
c_{k}=\sum_{j=1}^k (-1)^{[j]}\,, k=1,\ldots,\cM+\cN\,.
\label{eq:ck}
\end{equation}
Then, the dressing functions will keep the same form 
(\ref{eq:dress-ck}), but with now the above value for the parameters 
$c_{k}$.
Computing the residues for $\Lambda(u)$ with these new dressing 
functions, leads to the Bethe equations 
\begin{eqnarray}
&& 
(1-(-1)^{[l]}\langle \alpha_\ell, \alpha_\ell \rangle)\,
\prod_{k=1}^{\cM+\cN-1} \ 
\prod_{j=1}^{M^{(k)}}
\fe_{\langle\alpha_\ell,\alpha_k\rangle}(u_{i}^{(\ell)} -
u_{j}^{(k)}) 
= \frac{\lambda_{\ell}(u_i^{(\ell)}+\frac{\hbar}{2}\,c_{\ell})}
{\lambda_{\ell+1}(u_i^{(\ell)}+\frac{\hbar}{2}\,c_{\ell})}\ ,\nonu 
&& i=1,\ldots,M^{(\ell)}\,,\quad 1\leq\ell<\cM+\cN-1\,.
\qquad\qquad
\label{closedBAEgen}
\end{eqnarray} 
where $\langle\alpha_\ell,\alpha_k\rangle$ is the scalar product of
the simple roots, numbered 
\emph{as they are ordered by the chosen Dynkin diagram}. 
This single set of equations describe \textsl{all} the Bethe
equations, whatever the gradation (the Dynkin diagram) is, and
whatever the representations on each site of the spin chain are. 
In the particular case of only (mixture of) fundamental representation and/or its
contragredient on all sites, we
recover the isotropic limit ($q\to1$) of the spectrum and BAE computed in
\cite{RibMar}. These equations are also equivalent to the ones 
presented in \cite{zaka}, the different gradations here being related to 
the different possible paths (forms of the `hook') in \cite{zaka}.

Explicitely, in $sl(\cM|\cN)$, denoting $\alpha_{j}$ the simple roots, 
that we label according to 
their position $j=1,\ldots,\cM+\cN$ in the Dynkin diagram, their norm 
is given by $\langle\alpha_j,\alpha_j\rangle=(-1)^{[j]}\,2$ for 
the bosonic `white' roots  
and by $\langle\alpha_j,\alpha_j\rangle=0$ for the fermionic `grey' roots. 
Moreover, the scalar products between
different simple roots are all zero
\textsl{but} for the simple roots which are linked in the Dynkin 
diagram. 
Linked roots have scalar product 
$\langle\alpha_j,\alpha_{j+1}\rangle=-(-1)^{[j+1]}$. For 
more informations on the construction of simple roots and Dynkin
diagrams for superagebras, see e.g. \cite{dico}.

It should be clear that, since the different presentations (i.e. 
Dynkin diagrams) describe the same 
superalgebra and the same representations on the chain, 
the spectrum will be identical, although the dressing 
functions and the BAE look different. 

\subsection{Bethe equations for the symmetric grading}

In the case of $sl(\cM|2n)$, there exists a symmetric Dynkin diagram with two isotropic fermionic 
simple roots in positions $n$ and $\cM+n$:
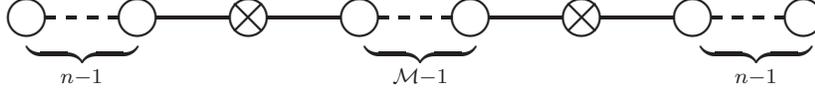
\begin{center}
\begin{picture}(452,20) \thicklines
   \multiput(80,10)(42,0){8}{\circle{14}}
   \put(159,5){\line(1,1){10}}\put(159,15){\line(1,-1){10}}
   \put(285,5){\line(1,1){10}}\put(285,15){\line(1,-1){10}}
   \put(87,10){\line(1,0){4}}\put(95,10){\line(1,0){4}}
   \put(103,10){\line(1,0){4}}\put(111,10){\line(1,0){4}}
   \put(129,10){\line(1,0){28}}
   \put(171,10){\line(1,0){28}}
   \put(213,10){\line(1,0){4}}\put(221,10){\line(1,0){4}}
   \put(229,10){\line(1,0){4}}\put(237,10){\line(1,0){4}}
   \put(255,10){\line(1,0){28}}
   \put(297,10){\line(1,0){28}}
   \put(339,10){\line(1,0){4}}\put(347,10){\line(1,0){4}}
   \put(355,10){\line(1,0){4}}\put(363,10){\line(1,0){4}}
   \put(80,0){$\underbrace{\hspace{42pt}}_{n-1}$}
   \put(208,0){$\underbrace{\hspace{42pt}}_{\cM-1}$}
   \put(335,0){$\underbrace{\hspace{42pt}}_{n-1}$}
\end{picture}
\end{center}
\vspace*{\baselineskip}
We give here the explicit expression for the dressing functions and Bethe Ansatz equations
for this diagram, thus taking $\cN = 2n$, and ordering the indices as in eq.(\ref{symmgrad}):
$$
[i]=\left\{ \begin{array}{ll}  0\,, & 1 \leq i \leq n \mb{and}
\cM+n+1\leq i\leq \cM+\cN\,,\\
                               1\,, & n+1\leq i \leq \cM+n\,.
  \end{array}\right.\,,
$$
i.e. 
\begin{equation}
\label{cksymm}
c_k = \left\{ \begin{array}{ll}  k\,, & k \leq n\,, \\
                                \cN- k\,, & n < k \leq \cM+n\,, \\
                                k -2\cM\,, & \cM+n < k \leq \cM+\cN\,. \end{array}\right.
\end{equation}
This choice of the grading implies that the elements of $T^{(\cM)}(u)$  (resp. $T^{(\cN)}(u)$) 
generate now a $\cY_{-\hbar}(\cM)$
(resp. $\cY_{\hbar}(\cN)$)
bosonic subalgebra. As a consequence, the expressions for the quantum Berezinian and its 
inverse are modified as follows:
$$
Ber(u) =  \mathrm{qdet}\, T^{(\cN)}(u-\hbar(\cM-\cN+1))\, 
\mathrm{qdet}\, T^{*(\cM)}(u-\hbar\cM)\,,
$$
$$
Ber^{-1}(u) = \mathrm{qdet}\, T^{*(\cN)}(u+\hbar(\cN-1))\, \mathrm{qdet}\, 
T^{(\cM)}(u-\hbar(\cM-\cN))\,.
$$
To determine its value on $v^+$ we rewrite the quantum Berezinian for the symmetric 
Dynkin diagram case as
\begin{eqnarray*}
Ber(u)&=& \sum_{\sigma \in S_{\cN}} \mathrm{sgn}\,(\sigma) 
T_{\sigma(1),1}(u-\hbar(\cM-\cN+1))\cdots
T_{\sigma(n),n}(u-\hbar(\cM-n)) \times \\
& \times & \, T_{\cM+\sigma(n+1),\cM+n+1}(u-\hbar(\cM-n+1))\cdots 
T_{\cM+\sigma(\cN),\cM+\cN}(u-\hbar\cM) \times
\\
& \times & \sum_{\tau \in S_{\cM}} \mathrm{sgn}\,(\tau) T^*_{n+\tau(1),n+1}(u-\hbar\cM)\cdots
T^*_{n+\tau(\cM),n+\cM}(u-\hbar)\,,
\end{eqnarray*}
obtaining:
$$
Ber(u)\,v^+  = \prod_{l=1}^n   \lambda_l(u-\hbar (\cM-l+1))
\prod_{l=n+1}^{\cM+n} \lambda^*_l(u-\hbar(\cM-l+n+1))
\prod_{l= \cM+n+1}^{\cM+\cN}\lambda_l(u-\hbar(2\cM-l+1))
\,v^+ 
$$
In the same way we can compute the constant value of $Ber^{-1}(u)$ on the $v^+$ module. 
Since
\begin{eqnarray*}
Ber^{-1}(u) &=&  \sum_{\sigma \in S_{\cN}} \mathrm{sgn}\,(\sigma) 
T^*_{\sigma(1),1}(u+\hbar(\cN-1))\cdots
T^*_{\sigma(n),n}(u+\hbar n) \times \\
& \times & \, T^*_{\cM+\sigma(n+1),\cM+n+1}(u+\hbar(n-1))\cdots 
T^*_{\cM+\sigma(\cN),\cM+\cN}(u) \times
\\
& \times & \sum_{\tau \in S_{\cM}} \mathrm{sgn}\,(\tau) 
T_{n+\tau(1),n+1}(u-\hbar(\cM-\cN))\cdots
T_{n+\tau(\cM),n+\cM}(u+\hbar(\cN-1))\,,
\end{eqnarray*}
we get
$$
Ber^{-1}(u)\,v^+  = \prod_{l=1}^n   \lambda^*_l(u+\hbar (\cN-l))
\prod_{l=n+1}^{\cM+n} \lambda_l(u-\hbar(\cN-l+n))
\prod_{l= \cM+n+1}^{\cM+\cN}\lambda^*_l(u+\hbar(\cM+\cN-l))
\,v^+\,.
$$
The steps leading to the dressing functions  (\ref{eq:dress-ck}) can now be repeated 
taking into account the different form of the value of the quantum Berezinian:
in particular, one can show that the constraints (\ref{fus3}) and (\ref{fus7}) are to be
replaced with: 
$$
A_0(u)\cdots A_{n-1}(u+\hbar(n-1))\,A^*_{n}(u)\cdots A^*_{\cM+n-1}(u+\hbar(\cM-1))\, 
\times 
$$
\begin{equation}
\times\, A_{\cM+n}(u+\hbar n)\cdots A_{\cM+\cN-1}
(u+\hbar(\cN-1)) = 1\,,
\end{equation}
and
$$
A^*_0(u+\hbar(\cN-1))\cdots A^*_{n-1}(u+\hbar n)\,A_{n}(u+\hbar(\cN-1))\cdots 
A_{\cM+n-1}(u+\hbar(\cN-\cM))\, \times 
$$
\begin{equation}
\times\, A^*_{\cM+n}(u+\hbar(n-1))\cdots A^*_{\cM+\cN-1}
(u) = 1\,.
\end{equation}
Both these constraints are satisfied by the dressing functions (\ref{eq:dress-ck}).
As a general rule, at each fermionic root two dressing functions 
$A$ and $A^*$ meet, and the $u^{(k)}_j$ parameters must satisfy an additional 
relation\footnote{In the distinguished
Dynkin diagram case there is only one fermionic root, corresponding to the 
$u^{*(\cM)}_j = u^{(\cM)}_j -\hbar \cM$ relation
obtained in the previous section.} 
of the form $u^{*(k)}_j = u^{(k)}_j -\hbar \cM$. 
We are now in position to write the Bethe Ansatz equations for the symmetric Dynkin 
diagram, requiring the transfer matrix  eigenvalue
$$
\Lambda(u) = \sum_{k=1}^{\cM+\cN}(-1)^{[k]} A_{k-1}(u) \lambda_k(u)
$$
to have vanishing residues at $u= u^{(l)}_j + \frac{\hbar}{2} c_l$ for $l=1,\dots,\cM+\cN-1$ 
and $j=1,\dots, M^{(l)}$.
The BAEs take the form:
\begin{eqnarray*}
&& \prod_{k=1}^{M^{(l-1)}}\fe_{-1}(u^{(l)}_j-u^{(l-1)}_k) 
\prod_{k \neq j}^{M^{(l)}} \fe_{2}(u^{(l)}_j-u^{(l)}_k)
\prod_{k=1}^{M^{(l+1)}}\fe_{-1}(u^{(l)}_j-u^{(l+1)}_k) =
\frac{\lambda_{l+1}(u_j^{(l)}+\frac{\hbar}{2}c_l)}
{\lambda_{l}(u_j^{(l)}+\frac{\hbar}{2}c_l)}\,, \\ 
&& \qquad 1\leq j\leq M^{(l)}\ ,\ 1\leq l<n \ \ \mathrm{and}\ \   
\cM+n+1<l < \cM+\cN 
\end{eqnarray*}
\begin{eqnarray*}
&& \prod_{k=1}^{M^{(n-1)}}\fe_{-1}(u^{(n)}_j-u^{(n-1)}_k) 
\prod_{k=1}^{M^{(n+1)}}\fe_{1}(u^{(n)}_j-u^{(n+1)}_k) = 
\frac{\lambda_{n+1}(u_j^{(n)}+\frac{\hbar}{2}n)}{\lambda_{n}(u_j^{(n)}+\frac{\hbar}{2}n)}\,, 
\end{eqnarray*}
\begin{eqnarray*}
&& \prod_{k=1}^{M^{(l-1)}}\fe_{1}(u^{(l)}_j-u^{(l-1)}_k) 
\prod_{k \neq j}^{M^{(l)}} \fe_{-2}(u^{(l)}_j-u^{(l)}_k)
\prod_{k=1}^{M^{(l+1)}}\fe_{1}(u^{(l)}_j-u^{(l+1)}_k) =
\frac{\lambda_{l+1}(u_j^{(l)}+\hbar(\cM-\frac{l}{2}))}
{\lambda_{l}(u_j^{(l)}+\hbar(\cM-\frac{l}{2}))}\,, \\
&& \qquad 1\leq j\leq M^{(l)}\ ,\ n<l<\cM+n 
\end{eqnarray*}
$$
\prod_{k=1}^{M^{(\cM+n-1)}}\fe_{1}(u^{(\cM+n)}_j-u^{(\cM+n-1)}_k) 
\prod_{k=1}^{M^{(\cM+n+1)}}
\fe_{-1}(u^{(\cM+n)}_j-u^{(\cM+n+1)}_k) = 
\frac{\lambda_{\cM+n+1}(u_j^{(\cM+n)}+
\frac{\hbar}{2}(n-\cM))}{\lambda_{\cM+n}(u_j^{(\cM+n)}+\frac{\hbar}{2}(n-\cM))}\,, 
$$
in agreement with eq.(\ref{closedBAEgen}).

\section{Open super-spin chains\label{sec:open}}

\subsection{Open chains monodromy and transfer matrices\label{monopen}}
As in the closed chain case
the supercommutation relations defining the reflection algebra allow us to show
that the transfer matrix
\begin{equation}\label{openb}
b(u) = str\left(K^{+}(u)\,B(u) \right) = \sum_{k,l=1}^{\cM+\cN}(-1)^{[k]}
K^{+}_{kl}(u)\,B_{lk}(u)\,.
\end{equation}
generates a commutative family
$$
\left[ b(u)\,,b(v)  \right] = 0\,,
$$
provided the matrix $K^+(u)$ obeys the `dual' reflection equation:
\begin{eqnarray}
&& R_{12}(-u+v)\ K_{1}^{+}(u)^{t}\
R_{21}(-u-v-\hbar(\cM-\cN))\ K_{2}^{+}(v)^{t}=
\nonumber \\
&&\qquad\qquad K^{+}_{2}(v)^{t}\ R_{12}(-u-v-\hbar(\cM-\cN))\ 
K_{1}^{+}(u)^{t}\ R_{21}(-u+v).\qquad
\label{red}
\end{eqnarray}
The classification of such matrices is deduced from the proposition 
\ref{reflectionclass}. Indeed, if $K^+(u)$ obeys the dual reflection
equation (\ref{red}), then 
$K^+(-u-\hbar\,\rho)^t$, with $\rho=\cM-\cN$, 
obeys reflection equation (\ref{reK}), so that 
$K^+(u) = U'\left(\EE' + \frac{\xi'}{u}\II \right)U^{'-1}$ for some 
new parameters $U'$, $\EE'$ and $\xi'$. 

We further assume that 
the matrix $K^+(u)$ commute with the matrix $K^-(v)$. Then, all the 
$K^pm(u)$ matrices are diagonalizable by the same matrix $U$, 
independent of the spectral parameter. Thus, one can assume that 
$K^{+}(u)$ is also diagonal and analytic:
\begin{equation}\label{Kplus}
K^{+}(u) =  \mathrm{diag}\,(\underbrace{\xi'-u,\dots, \xi'-u}_{L'_1}, 
\underbrace{u+\xi',\dots,u+\xi'}_{L'_2-L'_1},\xi'-u,\dots, \xi'-u)\,.
\end{equation}

Again, upon representation, one constructs a monodromy matrix
$\cB(u)$ for the $L$ sites open chain. 
%
In order to get analytical entries for the transfer matrix, we 
adopt the normalization (\ref{normaliz}) for $T(u)$ and $\mathcal T(u)$, 
and define:
\begin{eqnarray}
\cB(u) &=& \left(\cT(u)\otimes \cdots \otimes \cT(u) \right) K(u) 
\left(\cT^{-1}(-u)\otimes \cdots \otimes  \cT^{-1}(-u)\right)\,.\\
\wh b(u) &=& str \left( K^+(u)\, \cB(u) \right)
= \sum_{k=1}^{\cM+\cN}(-1)^{[k]}
K^{+}_{kk}(u)\,\cB_{kk}(u)\,.
\end{eqnarray}

\subsection{Symmetry of transfer matrices}

As we did in section \ref{invariance} for the closed chain case, we now turn to determine the 
 symmetry of the model whose transfer matrix is given by (\ref{openb}).
 Without any loss of generality 
 we assume in what follows that $L_1 < \cM<L_2$.
\begin{proposition}
We consider the transfer matrix $b(u)$ describing open spin chain models 
with boundary conditions given by $K(u)$ and $K^+(u)$, see eq.
(\ref{Kmat}) and (\ref{Kplus}), with $L_{1},L_{1}'<\cM$ and $L_2, L_{2}' > \cM$.
Let 
$$\fm_{j}=\min(L_{j},L_{j}')\mb{and}
\fM_{j}=\max(L_{j},L_{j}')\,,\qquad
j=1,2\,.$$
Then, $b(u)$
admits a $gl(\fm_1|\cM+\cN-\fm_2) \oplus \mathcal G \oplus gl(\cM-\fM_1|\fM_2-\cM)$ 
symmetry, where
$$
\mathcal G = \left\{  \begin{array}{l}   gl(\fM_{1}-\fm_1)\oplus gl(\fM_{2}-\fm_2)\,, \;\; if \;  
                                        (\fm_1\,,\fm_2) = (L_1\,,L_2) \; \,or \;
                                        (\fm_1\,,\fm_2) = (L'_1\,,L'_2)\,, \\ \\ 
                                        gl(\fM_{1}-\fm_1|\fM_{2}-\fm_2)\;\;\,otherwise\,.  \end{array} \right.
$$
\end{proposition}
\prf
 Supertracing in the first auxiliary space the supercommutation relations (\ref{recomponents}),
and expanding them in $u$ and $v$, one reads, from the $v^1$ order term
\begin{equation}\label{symmb}
\Big[ b(u)\,, B_{ij}^{(1)} \Big\} = -B_{ij}(u) \big(K^+_{ii}(u)-  
K^+_{jj}(u)\big) (\theta_i+\theta_j) \,.
\end{equation}
Since  $B_{ij}^{(1)}=0$ when $\theta_i +\theta_j=0$ (see eq. \ref{expB}),
one deduces that a non-zero generator
$B_{ij}^{(1)}$ commutes with $b(u)$ if and only if $ K^+_{ii}(u) =  
K^+_{jj}(u)$, that is to say $ \theta'_i = \theta'_j$. 
The symmetry (super)algebra is thus 
generated by the elements of $gl(L_{1}|\cM+\cN-L_{2})\oplus gl(\cM-L_{1}|L_{2}-\cM)$ 
obeying this relation: an enumeration of them 
 ends the proof.
\finprf

\subsection{Pseudovacuum for open chain transfer matrices}\label{Pv}
 A direct computation, using the result of 
propositions \ref{Brepr} and \ref{prop:Tinv}, shows that the highest weight vector $v^+$
is an eigenvector of $\wh b(u)$:
\begin{eqnarray*}
\wh b(u)\,v^+ &=& \sum_{j=1}^{\cM+\cN}(-1)^{[j]} K_{jj}^{+}(u)\, 
\cB_{jj}(u)\,v^+ = \Lambda_0(u) \,v^+\,,\\
\Lambda_0(u) &=& \sum_{j=1}^{\cM+\cN}(-1)^{[j]}\,{\wt g_j(u)}\, \beta_j(u)\,.
\end{eqnarray*}
The functions $\beta_{j}(u)$ are determined by the 
representations on the chain:
\begin{eqnarray}
\beta_j(u) &=& \left(\prod_{m=1}^{j-1}\lambda_m(-u + \hbar c_m )\right) \lambda_j(u)
\left(\prod_{m=j+1}^{\cM+\cN}\lambda_m(-u + \hbar c_{m-1})\right)\,.
\label{beta1}
\end{eqnarray}
In the above expressions the $\la_k(u)$'s are again the products of the 
eigenvalues for each site of the chain, as in (\ref{lambdarin}).

The functions $\wt g_j(u)$, $j=1,\dots,\cM+\cN$, depend only on the boundary 
matrices.  When $K^+(u)$ is the identity matrix, they take a simple form:
\begin{eqnarray}
\wt g_j(u) &=& \frac{2u (2u-\hbar\,c_{\cM+\cN})}{(2u-\hbar c_{j-1})(2u-\hbar{c_j})}\,
g_{j}(u)\label{gg}
\end{eqnarray}
where the functions $g_j(u)$ are defined in \eqref{gfunc}.

When $K^+(u)$ is of the form \eqref{diag}, they read:
\begin{eqnarray}
\wt g_j(u) &=& \frac{2u}{(2u-\hbar c_{j-1})(2u-\hbar{c_j})}\,g_{j}(u) \label{gg2}
\\
&&\times \begin{cases}  
\xi'\,(2u-\hbar\,c_{\cM+\cN})-u\big(2u-\hbar(2c_{L'_2}-2c_{L'_1}-c_{\cM+\cN})\big) &  \mbox{if } j<L'_1\\[1.2ex]
\xi'\,(2u-\hbar\,c_{\cM+\cN})+u\big(2u-\hbar(c_{\cM+\cN}-2c_{L'_2})\big) &  \mbox{if } L_1'\leq j<L'_2\\[1.2ex]
(\xi'-u)\,(2u-\hbar\,c_{\cM+\cN}) & \mbox{if } L'_2<j\end{cases}\nonumber
\end{eqnarray}
One recover the form \eqref{gg} by taking the limit $\xi'\to\infty$.

\subsection{Dressing functions for open chains}

We assume that all the eigenvalues of
$b(u)$ can be written as
\begin{equation}
\Lambda(u) = \sum_{k=1}^{\cM+\cN}\,(-1)^{[k]}\,\wt g_k(u)\,\beta_k(u)
\,A_{k-1}(u)\,,
\end{equation}
with $\wt g_k(u)$ and $\beta_k(u)$ given by (\ref{gg}) or \eqref{gg2}, 
and (\ref{beta1}) respectively, and dressing functions $A_{k}(u)$ to be determined.
The vanishing of the residues of $\Lambda(u)$ at $u= \frac{\hbar}{2}c_k$ 
implies that
$$
A_{k-1}(\frac{\hbar}{2}c_k) = A_k(\frac{\hbar}{2}c_k)\,, \qquad 
\mathrm{for}\quad 1\leq k \leq \cM-\cN-1\,.
$$
Using expression (\ref{hypdress}) for the dressing functions 
one can show that
the $M^{(k)}$'s are even and that
the simplest non--trivial way to satisfy the above constraint is to set
\begin{eqnarray*}
A_k(u) &=& \prod_{j=1}^{M^{(k)}} \frac{u-u_j^{(k)}-\frac{\hbar}{2} 
\left( c_{k+1}+(-1)^{[k+1]}\right)}
{u-u_j^{(k)}-\frac{\hbar}{2}\,c_k}\ 
\frac{u+u_j^{(k)}-\frac{\hbar}{2} \left( c_{k+1}+(-1)^{[k+1]}\right)}
{u+u_j^{(k)}-\frac{\hbar}{2}\,c_k} \\
&& \times \prod_{j=1}^{M^{(k+1)}} \frac{u-u_j^{(k+1)}-\frac{\hbar}{2} 
\left( c_{k}-(-1)^{[k+1]}\right)}
{u-u_j^{(k+1)}-\frac{\hbar}{2}\,c_{k+1}}\ 
\frac{u+u_j^{(k+1)}-\frac{\hbar}{2}\left( c_{k}-(-1)^{[k+1]}\right)}
{u+u_j^{(k+1)}-\frac{\hbar}{2}\,c_{k+1}}\,, 
\end{eqnarray*}
for $ k=0,\dots,\cM+\cN-1$, with the usual convention $M^{(0)} = M^{(\cM+\cN)} =0$.

\subsection{Bethe equations for the open chain}

In order to establish analyticity of all eigenvalues , one imposes
that the residues of $\Lambda(u)$ at $u= u^{(l)}_n + \frac{\hbar}{2}c_l$, 
for $1\leq n \leq M^{(l)}$, 
$ 0 < l \leq \cM+\cN-1$, all vanish. Using the definition (\ref{epsilon}) for the $\fe_n(u)$ function
one has the following set of Bethe Ansatz equation: 
\begin{eqnarray*}
&&\prod_{j \neq n}^{M^{(l)}} \fe_{2}(u^{(l)}_n-u^{(l)}_j )
\prod_{j =1}^{M^{(l)}} \fe_{2}(u^{(l)}_n+u^{(l)}_j )
\prod_{\tau = \pm 1}
\prod_{j =1}^{M^{(l+\tau)}} \fe_{-1}(u^{(l)}_n-u^{(l+\tau)}_j)
\,\fe_{-1}(u^{(l)}_n+u^{(l+\tau)}_j)=
\nonu
&&\qquad = \frac{\beta_l(u^{(l)}_n+\frac{\hbar}{2}c_l)}
{\beta_{l+1}(u^{(l)}_n+\frac{\hbar}{2}c_l)}
 \frac{\wt g_l(u^{(l)}_n+\frac{\hbar}{2}c_l)}{\wt g_{l+1}(u^{(l)}_n+\frac{\hbar}{2}c_l)}
\,, \qquad l=1\leq l<  \cM\,,\qquad\quad
\end{eqnarray*}
\begin{eqnarray*}
&& \prod_{j =1}^{M^{(\cM+1)}} \fe_{1}(u^{(\cM)}_n-u^{(\cM+1)}_j )
 \fe_{1}(u^{(\cM)}_n+u^{(\cM+1)}_j )
\prod_{j =1}^{M^{(\cM-1)}} \fe_{-1}(u^{(\cM)}_n-u^{(\cM-1)}_j)
\,\fe_{-1}(u^{(\cM)}_n+u^{(\cM-1)}_j)=\qquad
\nonu
&&\qquad = \frac{\beta_{\cM}(u^{(\cM)}_n+\frac{\hbar}{2}\cM)}
{\beta_{\cM+1}(u^{(\cM)}_n+\frac{\hbar}{2}\cM)}
\frac{\wt g_{\cM}(u^{(\cM)}_n+\frac{\hbar}{2}c_{\cM})}{\wt g_{\cM+1}(u^{(\cM)}_n
+\frac{\hbar}{2}c_{\cM})}\,, \qquad l=\cM\,,\qquad\quad
\end{eqnarray*}
\begin{eqnarray}
&& \prod_{j \neq n}^{M^{(l)}} \fe_{-2}(u^{(l)}_n-u^{(l)}_j )
\prod_{j =1}^{M^{(l)}} \fe_{-2}(u^{(l)}_n+u^{(l)}_j )
\prod_{\tau = \pm 1}
\prod_{j =1}^{M^{(l+\tau)}} \fe_{1}(u^{(l)}_n-u^{(l+\tau)}_j)
\,\fe_{1}(u^{(l)}_n+u^{(l+\tau)}_j)=
\nonu
&&\qquad = \frac{\beta_l(u^{(l)}_n+\frac{\hbar}{2}c_l)}
{\beta_{l+1}(u^{(l)}_n+\frac{\hbar}{2}c_l)}\,
\frac{\wt g_l(u^{(l)}_n+\frac{\hbar}{2}c_l)}{\wt g_{l+1}(u^{(l)}_n+\frac{\hbar}{2}c_l)}
\,, \qquad l=\cM< l < \cM+\cN\,.\qquad\quad   \label{BEEclosed}
\end{eqnarray}
As in the closed case, the left hand side of the Bethe equations only depends
on the choice of the algebra, while the right hand side explicitly depends on the choice
of the representation (through the $\beta_l(u)$ functions, eq. (\ref{beta1})) 
and on the reflection matrix 
(through the $\wt g_l(u)$ functions, eqs. (\ref{gg}) or \eqref{gg2}). 

\subsection{Bethe equations for other Dynkin diagrams}
We turn now to the calculation of the spectrum and Bethe equations of 
open super-spin chains for other Dynkin diagrams. The rules will be
the same as the ones given for the closed case (see section
\ref{sec:arbitraryDD}).
The functions $\wt g_{k}(u)$ have a form similar to
(\ref{gg}), with a change of increasing or decreasing
 behaviour of the poles  each
time a grey (fermionic) root is met, due to the change in the 
definition of the $\ZZ_{2}$-grading, and thus in the parameters 
$c_{k}$, as given in (\ref{eq:ck}).

The Bethe Ansatz equations read, for $\ell=1,\dots,\cM+\cN-1$ and
$i=1,\dots,M^{(\ell)}$ 
$$
\epsilon_{\ell} \,
\prod_{k=1}^{\cM+\cN-1} \ \prod_{j=1}^{M^{(k)}}
\fe_{\langle\alpha_\ell,\alpha_k\rangle}(u_{i}^{(\ell)} -
u_{j}^{(k)}) \ 
\fe_{\langle\alpha_\ell,\alpha_k\rangle}(u_{i}^{(\ell)} +
u_{j}^{(k)}) 
\ =\ \frac{\beta_{\ell}(u_i^{(\ell)}+
\frac{\hbar}{2}\,c_{\ell})}
{\beta_{\ell+1}(u_i^{(\ell)}+\frac{\hbar}{2}\,c_{\ell})}\,
\frac{\wt g_l(u^{(l)}_n+\frac{\hbar}{2}c_l)}{\wt g_{l+1}(u^{(l)}_i+\frac{\hbar}{2}c_l)}
\ ,\quad
$$
where $\epsilon_{\ell} = (1-(-1)^{[l]}\langle \alpha_\ell, \alpha_\ell \rangle)$, as
in the closed spin chain case.
As an example, we specialize the above formulas to  the symmetric Dynkin diagram case and $K^+(u)=\II_2$. 
The $\wt g$ functions are in this case:
\begin{eqnarray}
&& \wt g_l(u) = \frac{u (u + \frac{\hbar(\cM-\cN)}{2})}{(u+\frac{\hbar(l-1)}{2})(u+\frac{\hbar l}{2})}
\,,\quad l=1,\dots, \cN/2\,, \\
&&  \wt g_l(u) = \frac{u (u + \frac{\hbar(\cM-\cN)}{2})}{(u+\frac{\hbar(\cN-l+1)}{2})(u+\frac{\hbar(\cN-l)}{2})}\,,
\quad l=\cN/2+1,\dots, \cM+\cN/2\,, \\
&&  \wt g_l(u) = \frac{u(u + \frac{\hbar(\cM-\cN)}{2})}{(u+\frac{\hbar(l-2\cM-1)}{2})(u+\frac{\hbar(l-2\cM)}{2})}\,,
\quad l=\cM+\cN/2+1,\dots, \cM+\cN\,.
\end{eqnarray}
The Bethe equations, obtained by imposing analiticity of $\Lambda(u)$ at 
 points
$u= u^{(l)}_k+ \hbar c_l/2$, for $1 \leq k \leq M^{(l)}$ and 
$l=1,\dots,\cM+\cN-1$, are:
\begin{eqnarray}
&& \prod_{j \neq k}^{M^{(l)}}\fe_{2}(u^{(l)}_k-u^{(l)}_j)\prod_{j=1}^{M^{(l)}} \fe_{2} (u^{(l)}_k+u^{(l)}_j)
\prod_{\tau = \pm 1}\,\prod_{j=1}^{M^{(l+\tau)}} \fe_{-1}(u^{(l)}_k-u^{(l+\tau)}_j)\, \fe_{-1}(u^{(l)}_k+u^{(l+\tau)}_j)
= \nonu
&=& \frac{\beta_{l}(u_k^{(l)}+
\frac{\hbar}{2}\,c_{l})}
{\beta_{l+1}(u_k^{(l)}+\frac{\hbar}{2}\,c_{l})}\,
\frac{\wt g_l(u^{(l)}_n+\frac{\hbar}{2}c_l)}{\wt g_{l+1}(u^{(l)}_k+\frac{\hbar}{2}c_l)}\,,\quad 1 \leq l < n \ \mathrm{and} \ 
\cM+n < l < \cM+\cN\,,
\end{eqnarray}
\begin{eqnarray}
&& \prod_{j=1}^{M^{(n-1)}}\fe_{-1}(u_k^{(n)}-u_j^{(n-1)})\fe_{-1}(u_k^{(n)}+u_j^{(n-1)})
   \prod_{j=1}^{M^{(n+1)}}\fe_{1}(u_k^{(n)}-u_j^{(n+1)})\fe_{1}(u_k^{(n)}+u_j^{(n+1)}) = \nonu
&=& \frac{\beta_{n}(u_k^{(n)}+ \frac{\hbar}{2}\,c_{n})}
{\beta_{n+1}(u_k^{(n)}+\frac{\hbar}{2}\,c_{n})}\,
\frac{\wt g_n(u^{(n)}_k+\frac{\hbar}{2}c_n)}{\wt g_{n+1}(u^{(n)}_k+\frac{\hbar}{2}c_n)}\,, 
\qquad l=n\,,
\end{eqnarray}
\begin{eqnarray}
&& \prod_{j \neq k}^{M^{(l)}}\fe_{-2}(u^{(l)}_k-u^{(l)}_j)
\prod_{j=1}^{M^{(l)}} \fe_{-2} (u^{(l)}_k+u^{(l)}_j)
\prod_{\tau = \pm 1}\,
\prod_{j=1}^{M^{(l+\tau)}} \fe_{1}(u^{(l)}_k-u^{(l+\tau)}_j)\, \fe_{1}(u^{(l)}_k+u^{(l+\tau)}_j)
= \nonu
&=& \frac{\beta_{l}(u_k^{(l)}+
\frac{\hbar}{2}\,c_{l})}
{\beta_{l+1}(u_k^{(l)}+\frac{\hbar}{2}\,c_{l})}\,
\frac{\wt g_l(u^{(l)}_n+\frac{\hbar}{2}c_l)}{\wt g_{l+1}(u^{(l)}_k+\frac{\hbar}{2}c_l)}\,,
\qquad n < l < \cM+ n\,, 
\end{eqnarray}
\begin{eqnarray}
&& \prod_{j=1}^{M^{(l-1)}}\fe_{1}(u_k^{(l)}-u_j^{(l-1)})\fe_{1}(u_k^{(l)}+u_j^{(l-1)})
   \prod_{j=1}^{M^{(l+1)}}\fe_{-1}(u_k^{(l)}-u_j^{(l+1)})\fe_{-1}(u_k^{(l)}+u_j^{(l+1)}) = \nonu
&=& \frac{\beta_{l}(u_k^{(l)}+ \frac{\hbar}{2}\,c_{l})}
{\beta_{l+1}(u_k^{(l)}+\frac{\hbar}{2}\,c_{l})}\,
\frac{\wt g_l(u^{(l)}_k+\frac{\hbar}{2}c_l)}{\wt g_{l+1}(u^{(l)}_k+\frac{\hbar}{2}c_l)}\,, 
\qquad l=\cM+n\,. 
\end{eqnarray}

\section{Examples\label{sec:examples}}
In this section we discuss the application of our approach to few examples.
We will replace the $\hbar$ parameter with the imaginary unit $-i$, as it is customary in 
dealing with spin chains. 

Let us stress that, although in all examples, the energies will look identical (up to an 
irrelevant additive constant), the spectrum and 
Hamiltonians are indeed different. In fact, the 
energies are functions of the Bethe roots, which obey 
\underline{different} Bethe equations, specified by the
representations entering the spin chain.
\subsection{Closed super--spin chain in the fundamental representation}
Choosing for each site of the closed chain the fundamental representation,
we get the usual supersymmetric spin chains.
In the fundamental representation, one has $\mu^{[n]}_i = \delta_{i,1}$ 
for all sites $n=1,\dots,L$,
so that the eigenvalues (\ref{lambdarin}) become:
\begin{equation}\label{lambdafund}
\lambda_k(u) = \left\{ \begin{array}{ll}
\displaystyle\prod_{n=1}^L (u-a_n-i)\, & k=1 \,, \\[1.2ex]
\displaystyle\prod_{n=1}^L (u-a_n)\, & k\neq 1\,. \end{array}   \right.
\end{equation}
Plugging these expressions in the Bethe equations of section \ref{BE1}, we get
\begin{eqnarray}
&& \prod_{k=1}^{M^{(n-1)}}\fe_{-1}(u^{(n)}_j-u^{(n-1)}_k) 
\prod_{k \neq j}^{M^{(n)}} \fe_{2}(u^{(n)}_j-u^{(n)}_k)
\prod_{k=1}^{M^{(n+1)}}\fe_{-1}(u^{(n)}_j-u^{(n+1)}_k) = \nonumber \\
&&\qquad\qquad = \left\{  \begin{array}{ll} 
\displaystyle \prod_{l=1}^{L}\fe_{1}(u^{(1)}_j-a_l-i) & ,
\label{BEE0}\quad n=1 \\[1.4ex]
\displaystyle 1 & ,\quad 1 < n < \cM
\end{array}\right. \,,  \qquad 1\leq j\leq M^{(n)}\ , 
\end{eqnarray}
\begin{eqnarray}
&& \prod_{k=1}^{M^{(n-1)}}\fe_{1}(u^{(n)}_j-u^{(n-1)}_k) 
\prod_{k \neq j}^{M^{(n)}} \fe_{-2}(u^{(n)}_j-u^{(n)}_k)
\prod_{k=1}^{M^{(n+1)}}\fe_{1}(u^{(n)}_j-u^{(n+1)}_k) =
1\,,
\label{BEE1}\nonumber \\
&& 
\qquad\qquad 1\leq j\leq M^{(n)}\ ,\qquad \cM <n \leq \cM+\cN-1\,, 
\\[1.2ex]
&& \prod_{k=1}^{M^{(\cM-1)}}\fe_{-1}(u^{(\cM)}_j-u^{(\cM-1)}_k) 
\prod_{k=1}^{M^{(\cM+1)}}\fe_{1}(u^{(\cM)}_j-u^{(\cM+1)}_k) =
1\,,  \label{BEE2} \qquad 1\leq j\leq M^{(\cM)}\,.
\end{eqnarray}

Since here $T_{an}(u)=R_{an}(u)$, its value at $u=0$ is proportional 
to the graded permutation
operator between the $a$ (auxiliary) and $n$ (quantum) spaces. 
Thus, we can construct a local 
Hamiltonian in the usual way.
Choosing  $a_n = 0$ for all sites, we get
\begin{equation}
H = -i\,\frac{d}{du} \big(ln\; st(u)  \big)\Big|_{u=0}
= -\sum_{n=1}^{L} P_{n-1\,,n} \qquad \mathrm{with} \quad P_{01} = P_{L1}\,.
\end{equation}
Here $P_{n-1\,,n}$ is the graded permutation 
between sites $n-1$ and $n$. 
In particular, in the $\cM=1$, $\cN=2$  case we recover the supersymmetric 
$t-J$ model,
which corresponds to the $\cY(1|2)$ case \cite{Essler}.
The energies corresponding to the Hamiltonian (\ref{HAMFUN})
can be calculated by taking the logarithmic derivative of $\Lambda(u)$ and
evaluating at $u=0$, and are given by
$$
E= L + \sum_{j=1}^{M^{(1)}} \frac{1}{(u^{(1)}_j)^2+\frac{1}{4}}\,,
$$  
where the Bethe parameters $u_{j}^{(n)}$ are solution to the Bethe 
equations (\ref{BEE0}-\ref{BEE2}) with $a_{n}=0$, $\forall n$.\\

A slightly generalized case is obtained taking $a_p = a  \neq 0$ 
for a particular site $p$, and 
$a_n = 0$ for $n \neq p$.
This leads to the following Hamiltonian:
\begin{eqnarray*}
&& \label{HAMFUN} H = -\sum_{\stackrel{n=1}{n \neq p\,, p+1}}^L \!\!\!\! P_{n-1\,,n} + 
\frac{1}{a^2+1} 
\left( a^2 P_{p-1\,,p+1} + P_{p+1\,,p}   
 -i\, a P_{p+1\,,p-1} P_{p\,, p-1} + i\, a  P_{p\,,p-1}  P_{p+1\,,p-1} \right)\,.
\end{eqnarray*}
The energies get modified as follows:
$$
E= -L + \frac{a}{a+i}+ \sum_{j=1}^{M^{(1)}} \frac{1}{(u^{(1)}_j)^2+\frac{1}{4}}\,
$$
for the 
where the Bethe parameters $u_{j}^{(n)}$ are solution to the Bethe 
equations (\ref{BEE0}-\ref{BEE2}), with now inhomogeneities 
$a_{n}=\delta_{n,p}\,a$.

\subsection{Closed spin chain with an impurity}
Another case to which our formalism easily applies is the super--spin chain with one site
(the so--called impurity) in a representation different from the others. 
The easiest case is again the spin chain where all sites are in the fundamental representation
except for the $p^{th}$, associated to the highest weight $\mu_i^{[p]}$, $i=1,\dots,\cM+\cN$. 
The right hand sides of the Bethe equations are modified as follows:
\begin{equation}
\frac{\lambda_n(u^{(n)}_j-i\frac{n}{2})}{\lambda_{n+1}(u^{(n)}_j-i\frac{n}{2})}=
\left\{ \begin{array}{ll}\displaystyle
\Big(\fe_{1}(u^{(1)}_j -i) \Big)^{L-1}\, 
\frac{u^{(1)}_j-\frac{i}{2}-i \mu^{[p]}_1}{u^{(1)}_j-\frac{i}{2}-i \mu^{[p]}_{2}}
 & ,\,\, n=1\,, 
\\[2.1ex]
\displaystyle \frac{u^{(n)}_j-i\frac{n}{2}
 -i \mu^{[p]}_n}{u^{(n)}_j-i\frac{n}{2}-i \mu^{[p]}_{n+1}}
 & ,\,\, 1 <n< \cM\,,
\end{array}\right.
\end{equation}
\begin{equation}
\frac{\lambda_n(u^{(n)}_j-i\left(\cM-\frac{n}{2}\right))}{\lambda_{n+1}(u^{(n)}_j
-i\left(\cM-\frac{n}{2}\right))}=
\frac{u^{(n)}_j-i\left(\cM-\frac{n}{2}\right)-i \mu^{[p]}_n}{u^{(n)}_j
-i\left(\cM-\frac{n}{2}\right)-i \mu^{[p]}_{n+1}}\,,
\qquad \cM < n \leq \cM+\cN-1\,,
\end{equation}
\begin{equation}
\frac{\lambda_{\cM+1}(u^{(\cM)}_j-i\frac{\cM}{2})}{\lambda_{\cM}(u^{(\cM)}_j-i\frac{\cM}{2})} =
\frac{u^{(\cM)}_j-i\frac{\cM}{2}+i \mu^{[p]}_{\cM+1}}{u^{(\cM)}_j-i\frac{\cM}{2}-i\mu^{[p]}_{\cM}}\,,
\end{equation}
where we set again $a_n = 0$ for all sites.
The transfer  matrix and the  Hamiltonian of the $L$--sites spin chain with one impurity can 
be written as
\begin{eqnarray}
 st(u) &=& str_a \left(R_{a,1}(u)\cdots R_{a,p-1}(u)T_{a,p}(u)R_{a,p+1}(u)
\cdots R_{a,L}(u) \right)\,, \label{mon1imp} \\
 H &=&  -i\,T_{p+1,p}^{-1}(0)- 
P_{p-1\,,p+1}\, T_{p-1,p}^{-1}(0) \,
T_{p+1,p}(0)-\sum_{n=1\,,\,n\neq p-1,p}^L P_{n\,,n+1} \,. \label{H1imp}
\end{eqnarray}
It is worth noticing that all the quantum spaces $n$ (but the $p$--th one) 
are isomorphic to the auxiliary space $a$. Hence, $T_{n,p}(u)$, 
$n\neq p$, is defined in the same way $T_{a,p}(u)$ was introduced.
The spectrum of the Hamiltonian (\ref{H1imp}) is then given by:
$$
E= -(L-1) + i\, \frac{\mu'_1(0)}{\mu_1(0)} + \sum_{j=1}^{M^{(1)}} \frac{1}{(u^{(1)}_j)^2+\frac{1}{4}}\,.
$$

\subsection{Closed alternating spin chains}\label{Exampl}

In alternating spin chains, the spins along the chain belong alternatively to
two different representations.  As a particular example, one can take
an even number of sites $L$ for the chain, and let the spins in the even sites be
in the fundamental representation, while the spins in the odd sites are in a different one.
The transfer matrix  for such a chain will then be given by
$$
st(u) = str_a\,\left( T_{a,1}(u) R_{a,2}(u) \cdots  
T_{a,L-1}(u) R_{a,L}(u)   \right)\,,
$$
here the auxiliary space $a$ is $\cM+\cN$ dimensional.
One gets a local Hamiltonian
\begin{equation}\label{Alterham}
H=-i\,\left.\frac{d}{du} \left(ln\; st(u) \right)\right|_{u=0} = - 
\sum_{j =1}^{L/2}\, T^{-1}_{2j-2,2j-1}(0) 
\Big\{i\, \II+P_{2j-2,2j} T_{2j-2,2j-1}(0)\Big\}\,. 
\end{equation}
Denoting by $\mu_{j}$, $j=1,\ldots,\cM+\cN$ the weights of the 
representation on odd sites, and $\mu_{j}(u)= u -i\,(-1)^{[j]}\,\mu_{j}$,
 one gets for the eigenvalues (\ref{lambdarin})
$$
\lambda_k(u) = \left\{ \begin{array}{ll}
\big(u-i\big)^{L/2}\,\mu_{1}(u)^{L/2} & k=1\,, \\[1.4ex]
u^{L/2}\,\mu_{k}(u)^{L/2} \, & 1 < k \leq \cM+\cN\,. 
\end{array}   \right.
$$ 
where we set $a_n=0$ for all $n$. This leads to the spectrum
$$
E = -\frac{L}{2}\left(1-i\,\frac{\mu_{1}'(0)}{\mu_{1}(0)}\right)+
\sum_{j=1}^{M^{(1)}} \frac{1}{ (u^{(1)}_j)^2+\frac{1}{4}}\,,
$$

\subsubsection{Specialization to fundamental--adjoint alternating spin chain}
Choosing e.g. the adjoint representation for the odd sites, i.e.  highest weights  
$\mu^{[n]}_i = \delta_{i,1}$ 
for even $n$ and $\mu^{[n]}_i = \delta_{i1}+\delta_{i,\cM+\cN}$
for odd $n$, one gets the following form for the eigenvalues 
$$
\lambda_k(u) = \left\{ \begin{array}{ll}
\big(u-i\big)^{L} & k=1\,, \\[1.4ex]
u^L \, & 1 < k < \cM+\cN\,, \\[1.4ex]
\big(u+i\big)^{L/2}\,u^{L/2}\,,  & k =\cM+\cN\,,
\end{array}   \right.
$$ 
The  Bethe equations for $1 \leq n \leq \cM$ remain as in the fundamental representation case 
(\ref{BEE0}) and (\ref{BEE2}), while the equations (\ref{BEE1}) for $\cM < n \leq \cM+\cN-1$ 
are modified as follows:
\begin{eqnarray}
&& \prod_{k=1}^{M^{(n-1)}}\fe_{1}(u^{(n)}_j-u^{(n-1)}_k) 
\prod_{k \neq j}^{M^{(n)}} \fe_{-2}(u^{(n)}_j-u^{(n)}_k)
\prod_{k=1}^{M^{(n+1)}}\fe_{1}(u^{(n)}_j-u^{(n+1)}_k) = \nonumber \\
&& = \left\{  \begin{array}{ll} 
1   & ,\ \cM < n < \cM+\cN-1\,, \\[1.4ex]
\Big(\fe_{-1}(u^{(n)}_j-i \frac{\cM-\cN}{2}  )
 \Big)^{L/2}  & ,\ n=\cM+\cN-1 \,, 
\end{array}\right.     \label{Bethemod}
\end{eqnarray}
with $ 1\leq j\leq M^{(n)}$ . 
In this case, the monodromy matrix $T_{aj}(u)$
can  be obtained through the usual fusion procedure \cite{Fus1}, starting with the fused 
$R$ matrix: 
\begin{equation}\label{Rfus1}
R_{a(bc)}(u) = \mathcal P^+_{bc}R_{ac}(-u)R^{t_b}_{ab}(u)  \mathcal P^+_{bc}\,,
\end{equation} 
where $P^+_{bc} = \II_{bc}-\frac{1}{2\rho}Q_{bc}$ is a projector of dimension
$ \cM+\cN-1$. The tensor product of the spaces $b$ and $c$ is then considered as 
a single quantum space, and $T_{aj}(u)$ is 
obtained from $R_{a(bc)}$ through a suitable similarity transformation applied on both 
sides of (\ref{Rfus1}),
yielding:
\begin{eqnarray*} 
R_{aj}(u) &=& u \II_{aj}+i\, (\mathbf e_a \cdot \mathbf e_j)\,, \\
T_{aj}(u) &=& u \II_{aj}-i\, (\mathbf e_a \cdot \mathbf{\mathcal E}_j)\,,
\end{eqnarray*}
where $\mathbf e$ and $\mathbf{\mathcal E}$ respectively denote the $gl(\cM|\cN)$ generators
in the fundamental and adjoint representations. The inner product $\cdot$ is defined, as usual,
by means of the invariant, nondegenerate bilinear form $K^{\alpha \beta}$ 
on $gl(\cM|\cN)$, which is given as the supertrace on two generators
$ K_{\alpha \beta} = str\,\left( \mathcal E_{\alpha} \mathcal E_{\beta}\right)$:
$$
\mathbf A \cdot \mathbf B = \sum_{\alpha, \beta} (K^{-1})^{\alpha \beta} A_{\alpha}A_{\beta}
$$
Fusion allows also a direct calculation of $T_{aj}(u)^{-1}$, so that 
one gets an explicit expression of
the Hamiltonian (\ref{Alterham}). It involves nearest--neighbour and next--nearest--neighbour 
interaction terms:
\begin{equation}\label{Hamaltern}
H = \sum_{j=1, j\; even}^{L/2} H^{(1)}_{j,j+1} +  \sum_{j=1, j\; odd}^{L/2} H^{(2)}_{j-1,j,j+1}\,,
\end{equation}
where
\begin{eqnarray}\label{Hnn}
H^{(1)}_{j,j+1} &=& - \mathbf e_j \cdot \mathbf{\mathcal E}_{j+1} + \frac{1}{2\rho} 
\left( \mathbf e_j \cdot \mathbf{\mathcal E}_{j+1} \right)^2\,,
\qquad \rho = (\cM-\cN)/2\\
\label{Hnnn}
H^{(2)}_{j-1,j,j+1} &=& \frac{1}{2\rho}\left(\mathbf e_{j-1} \cdot {\mathcal E}_{j}\right)
\left\{ 2 \rho + \left( \mathbf e_{j-1} \cdot \mathbf{\mathcal E}_{j} \right) \right\}
(\mathbf e_{j-1} \cdot \mathbf e_{j+1})
\left( \mathbf e_{j-1} \cdot \mathbf{\mathcal E}_{j} \right)\,.
\end{eqnarray}
The spectrum of the Hamiltonian (\ref{Hamaltern}) reads:
$$
E = -L+\sum_{j=1}^{M^{(1)}} \frac{1}{ (u^{(1)}_j)^2+\frac{1}{4}}\,.
$$

\subsection{The open alternating spin chain}     
We define the transfer matrix for a $2L$--site open alternating chain as:
\begin{eqnarray*}
b(u) &=& str_a\Big(K^+(u)\,T_{a,1}(u)R_{a,2}(u)\cdots T_{a,2L-1}(u)R_{a,2L}(u)\,
K(u)\, \times\nonu
&&\qquad\times\,R^{-1}_{a,2L}(-u)T^{-1}_{a,2L-1}(-u)
\cdots R^{-1}_{a,2}(-u)T^{-1}_{a,1}(-u) \Big)
\end{eqnarray*}
Here the matrices acting on the even sites are in the fundamental representation, 
coinciding again with $R(u)$, and
the ones for the non--fundamental  are denoted with $T(u)$ and act on the odd sites 
of the chain. 
A local Hamiltonian can be obtained by taking the derivative of $b(u)$:
$$
H 
=\frac{1}{\xi\,\xi'\rho}\,\left.\frac{d}{du}b(u)\right|_{u=0}\,,
$$ 
where we remind $\rho=\cM-\cN$ while $\xi$ and $\xi'$ characterize the boundary 
matrices $K(u)$ and $K^+(u)$ respectively as in (\ref{Kmat}), (\ref{Kplus}).
One shows that
\begin{eqnarray*}
H &=& \frac{1}{\xi} K'_{2l}(0) + 
\frac{1}{\xi'\rho}\,str_a\left.\left(\frac{d\, K^+_a(u)}{du} \right)\right|_{u=0}
+\frac{2}{\rho}\,str_a\left\{\left( i\II+T_{a,1}(0)P_{a2} 
\right)T^{-1}_{a,1}(0)\right\} \\
&+& 2 \sum_{k=2}^l \left( i \II + T_{2k-2,2k-1}(0)P_{2k-2,2k} 
\right)T^{-1}_{2k-2,2k-1}(0)
\,.
\end{eqnarray*}
We will suppose that the gradation is such that $c_m \neq 0$ for $m > 0$.
In the case of distinguished gradation, this amounts to choose $\cM > \cN$.
Then, the energy spectrum is given by:
\begin{eqnarray}
E &=&  \beta\,L\,\left(1+\!\!\!\sum_{m=1}^{\cM+\cN-1} \frac{1}{c_{m}} 
-i\,\sum_{m=1}^{\cM+\cN} \frac{\mu'_{m}(-ic_{m-1})}{\mu_{m}(-ic_{m-1})} \right) 
- \sum_{j=1}^{M^{(1)}} \frac{2\,\beta}{(u^{(1)}_j)^2+\frac{1}{4}} \nonu
&& + i\,\beta\,\frac{\xi+\xi'}{\xi\,\xi'} +
2\,\beta\,\frac{1-\rho}{\rho}\,,
\end{eqnarray}
where  $\beta = \beta_1(0)$.
For the distinguished Dynkin diagram, and choosing the adjoint representation
for the odd sites, the Bethe equations read, for $1 \leq n \leq M^{(l)}$
with $1\leq l \leq \cM+\cN-1$:
\begin{eqnarray*}
&& \prod_{j=1}^{M^{(1)}} \fe_2(u_n^{(1)}-u_j^{(1)})\,  \fe_2(u_n^{(1)}+u_j^{(1)})\,
   \prod_{j=1}^{M^{(2)}} \fe_{-1}(u_n^{(1)}-u_j^{(2)})\,  \fe_{-1}(u_n^{(1)}+u_j^{(2)}) = \nonu
&&\qquad = - \Big( \fe_{-1}(u_n^{(1)}-i)\,\fe_{-3}(u_n^{(1)}+i) 
\Big)^L\ \fe_{1}(u^{(1)}_n)\ Q_1(u^{(1)}_n-\frac{i}{2})\,,
\end{eqnarray*}
\begin{eqnarray*}
&& \prod_{j = 1}^{M^{(l)}}\fe_{2}(u^{(l)}_k-u^{(l)}_j)\, \fe_{2} (u^{(l)}_k+u^{(l)}_j)
\prod_{\tau = \pm 1}\,\prod_{j=1}^{M^{(l+\tau)}} \fe_{-1}(u^{(l)}_k-u^{(l+\tau)}_j)\, 
\fe_{-1}(u^{(l)}_k+u^{(l+\tau)}_j) = \nonu
&&\qquad = -  \fe_{1}(u^{(l)}_n)\  Q_l(u^{(l)}_n-\frac{i\,l}{2})\,, \qquad 1 < l < \cM\,, 
\end{eqnarray*}
\begin{eqnarray*}
&& \hspace{-2.7ex}
\prod_{j =1}^{M^{(\cM+1)}} \fe_{1}(u^{(\cM)}_n-u^{(\cM+1)}_j )
 \fe_{1}(u^{(\cM)}_n+u^{(\cM+1)}_j )
\prod_{j =1}^{M^{(\cM-1)}} \fe_{-1}(u^{(\cM)}_n-u^{(\cM-1)}_j)
\,\fe_{-1}(u^{(\cM)}_n+u^{(\cM-1)}_j)=
\nonu
&&\qquad =  Q_{\cM}(u^{(\cM)}_n-\frac{i\,\cM}{2}) \,, 
\end{eqnarray*}
\begin{eqnarray*}
&& \prod_{j =1}^{M^{(l)}}\fe_{-2}(u^{(l)}_k-u^{(l)}_j)\, \fe_{-2} (u^{(l)}_k+u^{(l)}_j)
\prod_{\tau = \pm 1}\,\prod_{j=1}^{M^{(l+\tau)}} \fe_{1}(u^{(l)}_k-u^{(l+\tau)}_j)\, 
\fe_{1}(u^{(l)}_k+u^{(l+\tau)}_j) = \nonu
&&\qquad = -  \fe_{-1}(u^{(l)}_n)\ Q_l(u^{(l)}_n-\frac{i}{2}(2\cM-l))\,, 
\qquad \cM < l < \cM+\cN-1\,,
\end{eqnarray*}
\begin{eqnarray*}
&& \prod_{j=1}^{M^{(l)}}\fe_{-2}(u^{(l)}_k-u^{(l)}_j)\, \fe_{-2} (u^{(l)}_k+u^{(l-1)}_j)
\prod_{\tau = \pm 1}\,\prod_{j=1}^{M^{(l+\tau)}} \fe_{1}(u^{(l)}_k-u^{(l+\tau)}_j)\, 
\fe_{1}(u^{(l)}_k+u^{(l+\tau)}_j)=
\nonu
&&\qquad = - \Big( \fe_{-1}(u_n^{(l)}-i\,\rho)\,  \fe_{-1}(u_n^{(l)}+
i\,\rho)   \Big)^L\, \fe_{1}(u^{(l)}_n)\  
Q_{\cM+\cN-1}\big(u^{(l)}_n-i(\rho-\frac{1}{2})\big)\,, \nonu
&&  l = \cM+\cN-1\,.\nonumber
\end{eqnarray*}
In the above equations, we set 
$$
Q_l(u) = \frac{\wt g_l(u)}{\wt g_{l+1}(u)}\,,
$$ 
according to the chosen boundary matrices, see eqs. \eqref{gg} and \eqref{gg2}.

\section*{Acknowledgments} 
We are grateful to R. Nepomechie who pointed out some mistakes in the expression of the vacuum eigenvalue for the open case.


\end{document}